\documentclass[reprint,amsmath,amssymb,prl]{revtex4-1}

\usepackage{color}
\usepackage{graphicx}
\usepackage{bm}
\usepackage{amsmath,amssymb}
\usepackage{xspace}
\usepackage{braket}

\AtBeginDocument{\let\oldcontentsline\contentsline}
\newcommand{\notoccontentsline}[4]{\oldcontentsline{}{}{}{}}
\newcommand{\droptocpage}{\addtocontents{toc}{\let\protect\contentsline\protect\notoccontentsline}}
\newcommand{\incltocpage}{\addtocontents{toc}{\let\protect\contentsline\protect\oldcontentsline}}

\begin{document}
	\title{Observation of Laughlin states made of light}

	\author{Logan W. Clark$^*$}
	\author{Nathan Schine$^*$}
	\author{Claire Baum}
	\author{Ningyuan Jia}
	\author{Jonathan Simon}
	\affiliation{James Franck Institute and Department of Physics, University of Chicago, Chicago, IL 60637, USA}
	    
\date{\today}

	\begin{abstract}
		Much of the richness in nature emerges because the same simple constituents can form an endless variety of ordered states~\cite{Anderson2019}. While many such states are fully characterized by their symmetries~\cite{LandauBook}, interacting quantum systems can also exhibit topological order, which is instead characterized by intricate patterns of entanglement~\cite{Chen2010, Nayak2008}.
		A paradigmatic example of such topological order is the Laughlin state~\cite{Laughlin1983}, which minimizes the interaction energy of charged particles in a magnetic field and underlies the fractional quantum Hall effect~\cite{Tsui1982}. Broad efforts have arisen to enhance our understanding of these orders by forming Laughlin states in synthetic quantum systems, such as those composed of ultracold atoms~\cite{Bloch2008, Cooper2019} or photons~\cite{Umucalilar2014, Carusotto2013, Ozawa2019}. In spite of these efforts, electron gases remain essentially the only physical system in which topological order has appeared~\cite{Tsui1982, Du2009, Bolotin2009, Spanton2018}.
		Here, we present the first observation of optical photon pairs in the Laughlin state. 
		These pairs emerge from a photonic analog of a fractional quantum Hall system, which combines strong, Rydberg-mediated interactions between photons~\cite{Jia2018b, Peyronel2012, Birnbaum2005, Thompson2013} and synthetic magnetic fields for light, induced by twisting an optical resonator~\cite{Schine2016, Clark2019, Ozawa2019}.
		Photons entering this system undergo collisions to form pairs in an angular momentum superposition consistent with the Laughlin state. Characterizing the same pairs in real space reveals that the photons avoid each other, a hallmark of the Laughlin state.
		This work heralds a new era of quantum many-body optics, where strongly interacting and topological photons enable exploration of quantum matter with wholly new properties and unique probes. 
	\end{abstract}

	\maketitle

	The simplest recipe for realizing topologically ordered many-body states is to place strongly-interacting particles in an effective magnetic field. The magnetic field quenches the kinetic energy of the particles, so that they order themselves solely to minimize their interaction energy, forming intricate patterns of long-range entanglement~\cite{Chen2010}.
	These states exhibit fascinating properties largely unseen in other forms of matter; for example, in addition to the robust quantized edge transport which also appears in weakly-interacting systems~\cite{Hasan2010}, topologically ordered phases can host excitations with fractional charge and anyonic exchange statistics~\cite{Stern2008}. More exotic phases can even host non-Abelian anyons, a promising constituent for fault-tolerant quantum computers thanks to their insensitivity to local perturbations~\cite{Nayak2008}.

	\begin{figure*}
		\centering
		\includegraphics{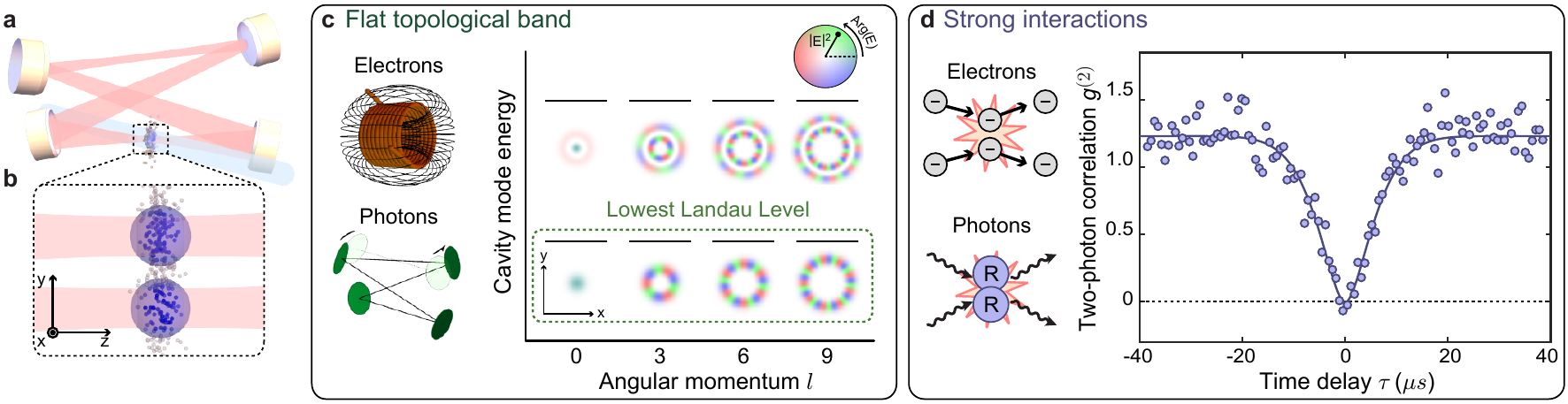} 
		\caption{
			\textbf{The ingredients for topologically ordered light.}
			\textbf{a,} Our experiment couples optical photons (red) with a gas of Rubidium atoms at the waist of a twisted, four mirror optical cavity. \textbf{b,} This coupling turns each photon entering the cavity into a \textit{polariton}, a quasiparticle combining the photon with a collective Rydberg excitation of the atomic gas. Polaritons can move around in the transverse modes available to their photonic component, and a pair of polaritons (depicted) strongly interacts because of their Rydberg components (blue spheres). 
			\textbf{c,} Two key ingredients enable this system to explore topological order. First, we form a flat topological band of single-photon states using a twisted optical cavity, which hosts a set of degenerate photonic modes that are equivalent to the states in the lowest Landau level available to electrons in a strong magnetic field. 
			\textbf{d,} Second, the strong polaritonic interactions are analogous to the Coulomb interactions between electrons in a traditional fractional quantum Hall system. 
			Polaritons confined to a single cavity mode reveal their strong interactions via transport blockade, wherein a single polariton present in the cavity prevents a second photon from entering. Blockade results in antibunched correlations of photons exiting the cavity, shown here for $l=0$.
			\label{fig:Intro}}
	\end{figure*}

	Few experimental systems have been found to host topologically ordered states. All definitive observations of such order have been made in two-dimensional electron gases subjected to magnetic fields, originally in semiconductor heterojunctions~\cite{Tsui1982}, as well as more recently in graphene~\cite{Du2009, Bolotin2009} and van der Waals bilayers~\cite{Spanton2018}.

	The scarcity of physical platforms hosting topological order has spurred great interest in elucidating its exotic properties using the wide tunability, particle-resolved control, and versatile detection capabilities afforded by synthetic quantum matter~\cite{Bloch2008, Carusotto2013,Ozawa2019,Cooper2019}. The constituents of typical synthetic matter are atoms and photons, which do not experience a Lorentz force in ordinary magnetic fields because they are charge neutral. Therefore, the key challenge is to implement a synthetic magnetic field which creates an effective Lorentz force and is compatible with strong interactions between particles. A classic approach employed the Coriolis force in rotating ultracold atomic gases~\cite{Cooper2008}, and such systems approached the few-body fractional quantum Hall regime~\cite{Gemelke2010}. More recent efforts in ultracold atoms focused on Floquet engineering of synthetic magnetic fields~\cite{Eckardt2017} combined with strong atomic interactions thanks to tight confinement in an optical lattice~\cite{Tai2017}. Photonic systems have also demonstrated a variety of synthetic magnetic fields~\cite{Ozawa2019} compatible with strong interactions via coupling to superconducting qubits in the microwave domain~\cite{Wallraff2004, Roushan2017} and cold atoms~\cite{Birnbaum2005,Peyronel2012,Thompson2013,Jia2018b} or quantum dots~\cite{barik2018topological} in the optical domain. Because these ingredients have yet to be effectively combined and scaled, the formation of topologically ordered synthetic matter has remained elusive.

	In this work we observe the formation of optical photon pairs in a Laughlin state. To this end, we construct a photonic system analogous to an electronic fractional quantum Hall fluid by combining two key ingredients: a synthetic magnetic field for light induced by a twisted optical cavity~\cite{Schine2016} and strong photonic interactions mediated by Rydberg atoms~\cite{Jia2018b}. We first observe that photons in this system undergo collisions which satisfy conservation laws and have density-dependence characteristic of two-body processes. Correlations in the resulting photon pairs reveal a two-photon angular momentum distribution consistent with a Laughlin state. Moreover, characterizing these photon pairs in real space reveals that they strongly avoid being in the same location. Together, these results indicate that the photon pairs have $76(18)\%$ overlap with a pure Laughlin state.

	Our typical experiment begins by loading a gas of 6000(1000) laser-cooled Rubidium-87 atoms at the waist of an optical cavity (Fig.~\ref{fig:Intro}a; Methods~\ref{methods:ExperimentSetup}). We then continuously shine a weak probe laser beam on the cavity for $100$~ms. The initially uncorrelated photons from the probe laser which enter the cavity are strongly coupled to a resonant atomic transition; in the presence of an additional Rydberg coupling field the photons are converted into polaritons, quasi-particles consisting of a superposition between a photon and a collective Rydberg excitation of the atomic gas (Fig.~\ref{fig:Intro}b)~\cite{Peyronel2012,Jia2018b}. Photons emerge on the other side of the cavity in a state which is highly correlated in both real space and angular momentum space, whose structure reflects the steady state formed by the intracavity polaritons.

	Key properties of the polaritons, inherited from both their photonic and Rydberg components, make them behave like particles in a fractional quantum Hall system. First, the motion of individual polaritons is determined by the cavity modes accessible to their photonic part~\cite{Sommer2016}. The large energy spacing between longitudinal mode manifolds restricts the polaritons to a single manifold, confining them to undergo two-dimensional motion among the transverse modes. We utilize a twisted optical cavity whose transverse mode spectrum forms sets of degenerate orbital angular momentum states that are equivalent to the Landau levels accessible to electrons in a magnetic field~\cite{Schine2016} (Fig.~\ref{fig:Intro}c). 
	Thus, we create a synthetic magnetic field for polaritons, realizing the first key ingredient of topological order.

	To form ordered states, polaritons must also interact with one another (Fig.~\ref{fig:Intro}d). Polaritons in our system inherit the strong interactions of their Rydberg components~\cite{Saffman2010}, causing them to avoid each other.
	When confined to a single transverse mode, one polariton can only avoid another by blockading the cavity, preventing a second polariton from entering~\cite{Jia2018b}. This phenomenon can be characterized via the two-photon correlation function $g^{(2)}(\tau)$, which quantifies the likelihood of seeing two photons emerge from the cavity separated by a time $\tau$ compared to completely uncorrelated photons (Methods~\ref{methods:CorrelationsG2}). Blockade manifests strikingly through antibunching; the same-time correlation $g^{(2)}(0)$ falls to zero because there are never two polaritons present in the cavity at the same time.

\begin{figure}
		\centering
		\includegraphics[width=0.95\columnwidth]{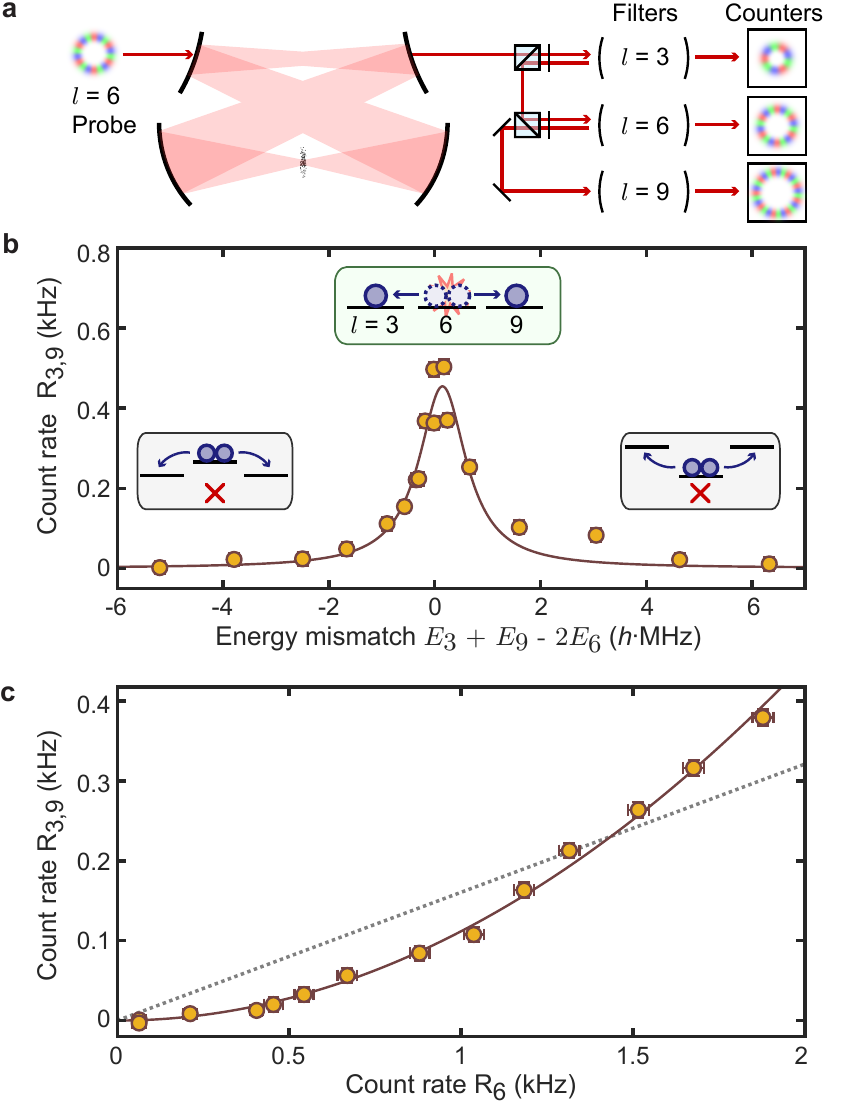}
		\caption{
			\textbf{Collisions between polaritons in the lowest Landau level.}
			\textbf{a,} To test for mode-changing collisions between polaritons, we inject photons with orbital angular momentum $l=6$ and then count the photons which emerge from the cavity in each angular momentum state (Methods~\ref{methods:ModeSorting}).
			\textbf{b,} The total rate $R_{3,9}$ at which photons emerge from the cavity with angular momentum $l=3$~or~$9$ exhibits a peak when the collision process depicted in the insets conserves energy; $E_l$ is the energy of a polariton with angular momentum $l$. The data are well fit by a lorentzian (solid curve).
			\textbf{c,} When all three states are degenerate and we vary the power of the probe laser, the count rate $R_{3,9}$ grows quadratically (solid curve) rather than linearly (dotted line) with the count rate $R_6$ of $l=6$ photons, consistent with the production of $l=3~\&~9$ polaritons in collisions between two $l=6$ polaritons. 
			All error bars indicate standard error.
			\label{fig:ModeChangingCollisions}}
	\end{figure}

	When polaritons have access to multiple transverse modes, in our case in the lowest Landau level, new physics emerges: it becomes possible for two polaritons to enter in the cavity at the same time while still avoiding one another. In this case, interactions need not lead to blockade, but can instead drive collisions between polaritons which cause them to move among the states of the Landau level. To test for these collisions, we provide the polaritons access to exactly three states in the lowest Landau level -- those with orbital angular momentum $l/\hbar=3,6,$ and $9$ -- using a Floquet scheme~\cite{Clark2019} (Methods~\ref{methods:FloquetScheme}). Next, utilizing a digital micromirror device for wavefront shaping, we shine a laser on the cavity which only injects photons with angular momentum $l=6$. After the photons have interacted with one another as polaritons in the cavity, they emerge from the other side, where we sort them and count how many emerge in each angular momentum state (Fig.~\ref{fig:ModeChangingCollisions}a).

	Despite the exotic nature of polaritonic quasiparticles, we find that polaritons undergo collisions much like ordinary particles. In particular, collisions between polaritons must conserve the total energy, as well as angular momentum thanks to the rotational symmetry of the system. Accordingly, when we inject photons into the cavity with angular momentum $l=6$ (as we will do throughout this work), the only collision process which conserves angular momentum converts two input polaritons with $l=6$ into one output polariton with $l=3$ and another with $l=9$. Similarly, as we tune the relative energies between the different angular momentum states, we only observe photons emerging with $l=3$ or $9$ when the aforementioned collision process can conserve energy (Fig.~\ref{fig:ModeChangingCollisions}b). 
	
	Multiple particles must encounter one another in order to collide. This requirement generically manifests in a collision rate which is nonlinear in the density of particles present in the system, much like reaction rates in chemistry. Indeed, we find that increasing the probe beam power raises the density of polaritons and thereby makes them more likely to collide (Fig.~\ref{fig:ModeChangingCollisions}c). 
    In particular, the rate at which collision products appear $R_{3,9}\propto R_6^2$ is quadratic in the rate $R_6$ at which photons emerge in the initial angular momentum state. Because of the strong Rydberg-mediated interactions, we observe these collisions even though there are never more than two polaritons present simultaneously.

	\begin{figure*}
		\centering
		\includegraphics[width=0.9\textwidth]{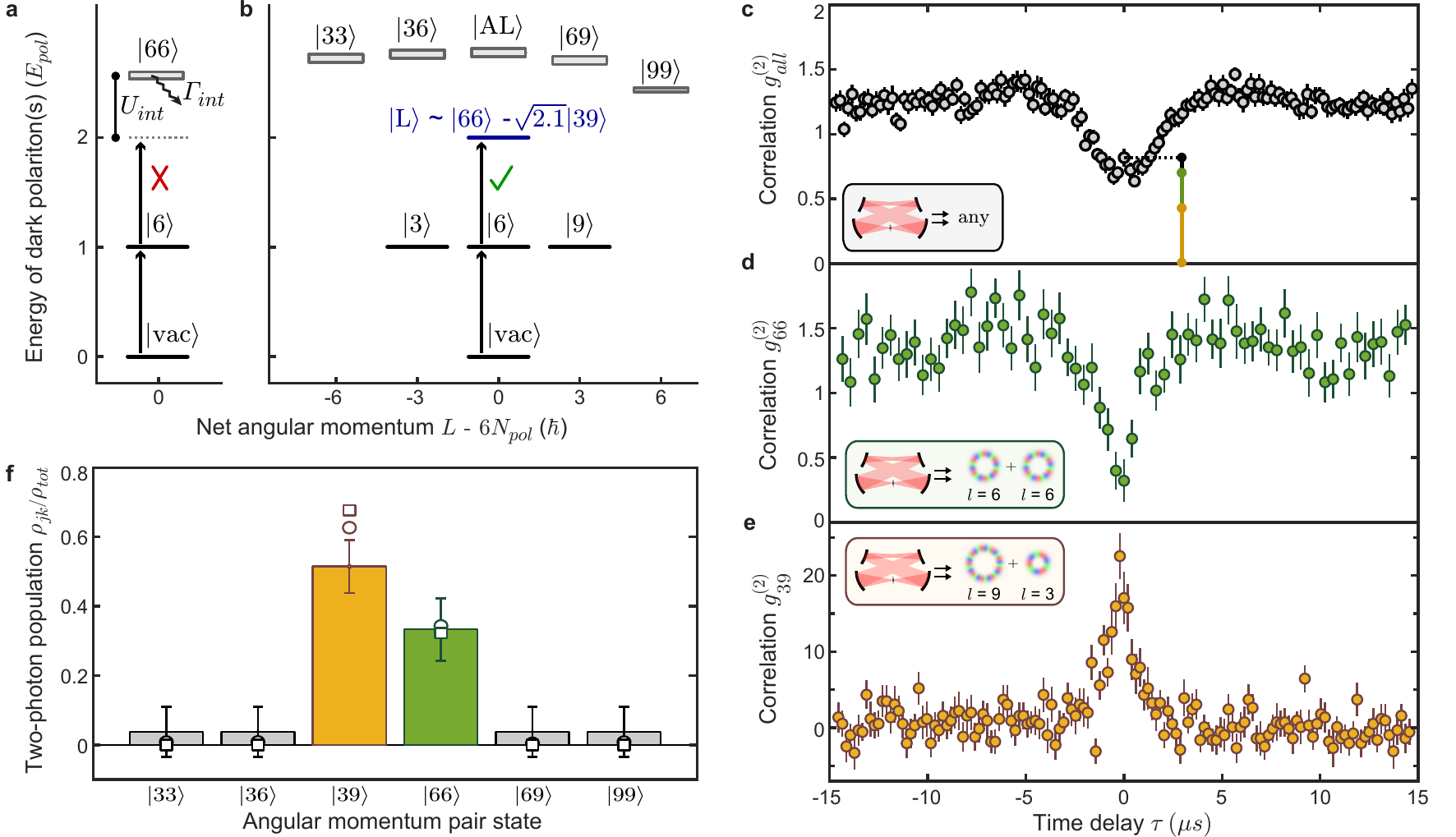}
		\caption{
			\textbf{Laughlin state characterization in angular momentum space.}
			Schematic of the many-body energy levels for states containing polariton numbers $N_{pol}\leq2$ exploring a single mode (\textbf{a}) or all three modes of the lowest Landau level (\textbf{b}).
			In a single mode (here $l=6$), the two polariton state $\ket{66}$ has its energy shifted ($U_{int}$) and rapidly decays ($\Gamma_{int}$) due to interactions; these effects induce blockade. 
			With three modes accessible, most two-polariton states are still shifted by interactions. However, a Laughlin state $\ket{L}$ arises in which the interference between $\ket{66}$ and $\ket{39}$ suppresses the interaction energy (to zero in the limit of contact interactions). This pair state remains long-lived and can be resonantly excited by the $l=6$ probe laser (black arrows). 
			\textbf{c,} When all emitted photons are included regardless of angular momentum, photons with access to all three modes exhibit weak blockade.
			\textbf{d,} Correlations between photons exiting the cavity in the probed angular momentum mode ($l=6$) exhibit stronger blockade, with remnant nonzero correlation $g_{66}(0)=0.32(3)$ at zero delay. \textbf{e,} Correlations between photons with angular momenta $l=3$ and $l=9$ exhibit a large positive correlation because they are produced together in collisions.
			\textbf{f,} Relative populations in the two-photon manifold determined from coincidence events (bars) are comparable to the Laughlin state (squares) and an atomistic numerical model \cite{Clark2019} (circles). The contributions of each pair state to $g^{(2)}_{all}(0)$ are represented by the colored, vertical bars in panel (c), see Methods~\ref{methods:AngMomCorr} for details.  All error bars indicate standard error. 
			\label{fig:AngularMomentumSpace}}
	\end{figure*}

	Ordered states emerge as a result of these collisions between polaritons. To better understand the ordered states accessible to polaritons in our system, we consider the energy spectrum of zero, one, and two polariton states in Fig.~\ref{fig:AngularMomentumSpace}a \& b. We focus on the case in which all of the single-polariton states $\ket{l}$ are degenerate with energy $E_{pol}$. When only a single cavity mode, for instance $l=6$, is accessible, the interactions between polaritons cause the state $\ket{66}$ with two polaritons in that mode to have a shifted energy and a much shorter lifetime than it would without interactions. Thus, a probe laser which resonantly excites $\ket{6}$ from the vacuum state $\ket{vac}$ does not subsequently excite the $\ket{6}\rightarrow\ket{66}$ transition, leading to blockade like that shown in Fig.~\ref{fig:Intro}f. Such blockade precludes the formation of multi-photon states.
	
	When three modes in the lowest Landau level are accessible, a long-lived Laughlin state $\ket{L}$ emerges in the two-particle sector (Fig.~\ref{fig:AngularMomentumSpace}b). Our experiments offer a unique opportunity to connect the mathematical form of this Laughlin state to observations of its microscopic structure. For the particular modes used in this work, the two-particle Laughlin wavefunction in real space is $\psi_L(z_1, z_2)\propto z_1^3z_2^3(z_1^3-z_2^3)^2\exp(-|z_1|^2/4-|z_2|^2/4)$ where $z_k\equiv x_k+iy_k$ is a complex number reflecting the position $(x_k,y_k)$ of particle $k$. Expanding the quadratic factor makes it possible to write this state in angular-momentum space as $\ket{L}=\frac{1}{\sqrt{3.1}}\ket{66}-\sqrt{\frac{2.1}{3.1}}\ket{39}$, where $\ket{mn}$ is the state with two polaritons of angular momenta $m\hbar$ and $n\hbar$ (SI~\ref{SI:VarietiesOfLaughlinStates}). Because the wavefunction goes to zero when the particles are at the same position ($\psi_L(z_1=z_2)=0$), residing in the Laughlin state enables two particles to avoid each other while remaining in the lowest Landau level. From the perspective of angular momentum states, this avoidance arises from destructive interference between the $\ket{66}$ and $\ket{39}$ two-photon amplitudes for particles at the same location. Similar two-particle Laughlin states can be formed among any set of three modes with evenly spaced angular momenta (see SI~\ref{SI:VarietiesOfLaughlinStates}).
	
	The spatial anticorrelation of photons in the Laughlin state suppresses the interaction energy shift and interaction-induced decay experienced by other two-particle states. Thus, when we shine an ordinary laser into this atom-cavity system, we anticipate that a two-photon Laughlin state will emerge from the other side, because all other two-particle states are blockaded (including the ``Anti-Laughlin'' superposition state $\ket{AL}=\sqrt{\frac{2.1}{3.1}}\ket{66}+\frac{1}{\sqrt{3.1}}\ket{39}$). From this perspective, the collisions observed in Fig.~\ref{fig:ModeChangingCollisions} are in fact the first hint of the formation of Laughlin states.

	To definitively test for the formation of Laughlin states we characterize the correlations of the photon pairs emerging from the cavity~\cite{Umucalilar2014}. First, when we detect all output photons regardless of their spatial mode, the ``all-mode'' correlations $g^{(2)}_{all}(\tau)$ reveal only a weak blockade effect and in fact have a local maximum at time zero (Fig.~\ref{fig:AngularMomentumSpace}c). 
	This weak blockade confirms that photon pairs are now able to traverse the cavity, but determining their structure requires more detailed measurements.

	\begin{figure}
		\centering
		\includegraphics{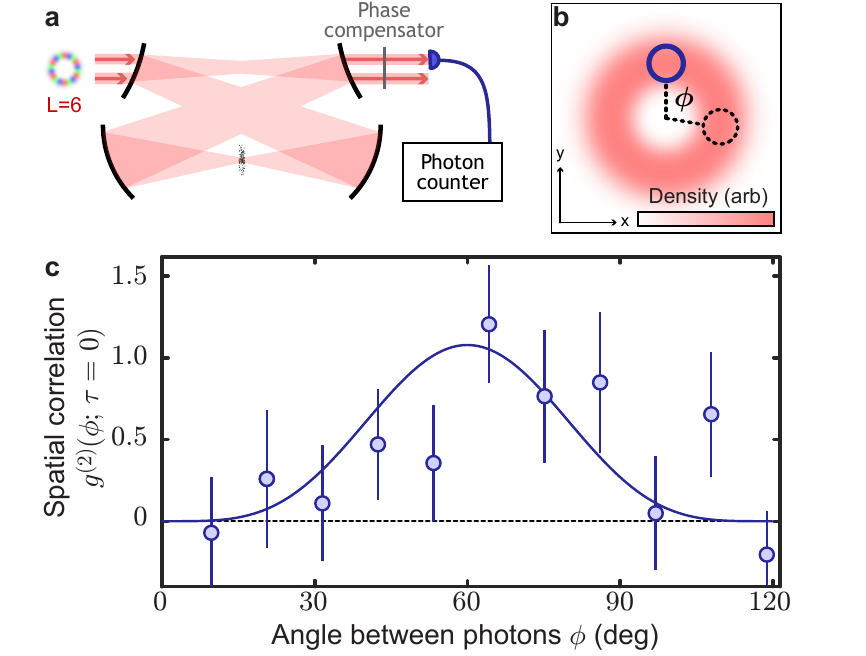} 
		\caption{
			\textbf{Spatial correlations of a photonic Laughlin puddle.}
			\textbf{a,} A single-mode fiber (purple) admits only photons at its location, enabling us to filter photons by their real-space position. A mode-dependent phase compensator counteracts the phase difference between polaritons and photons imprinted by our Floquet scheme (Methods~\ref{methods:SpaceCorr}). \textbf{b,} The average density of photons forms a smooth annulus with no angular structure. We place the fiber off-center (solid circle) at the radius with the highest density and measure correlations as a function of the angle $\phi$ to a second location (dashed circle) around the annulus. \textbf{c,} The measured angular correlations (circles) for zero time delay exhibit a periodic structure consistent with the expected form $g_L^{(2)}(\phi)\propto\sin^4(3\phi/2)$ for a Laughlin state (solid curve). The minimum of correlation near $\phi=0$ reveals that photons avoid being in the same location. Error bars indicate standard error.
			\label{fig:SpatialCorrelation}
		}
	\end{figure}
	
	We gain deeper insight by examining the correlations $g^{(2)}_{jk}$ between photons with angular momenta $l=j$ and $l=k$, again using the setup shown in Fig.~\ref{fig:ModeChangingCollisions}a. The correlations $g^{(2)}_{66}$ between photons with $l=6$ have a nonzero value $g^{(2)}_{66}(0)=0.32(3)$ at zero time delay, indicating substantial population in the pair state $\ket{66}$ (Fig.~\ref{fig:AngularMomentumSpace}d). However, their blockade is still much deeper than that of the all-mode correlations $g^{(2)}_{all}$, indicating that $g^{(2)}_{all}$ has a large contribution from other pairs which are not in the state $\ket{66}$. Most of the remaining pairs are accounted for by examining $g^{(2)}_{39}$, which exhibits a prominent peak at time zero (Fig.~\ref{fig:AngularMomentumSpace}e). The peak height indicates that photons are $g^{(2)}_{39}(0)=22(2)$ times more likely to appear in both modes simultaneously than we would expect for uncorrelated photons arriving with the same individual rates. This bunching arises because photons in these modes are predominantly produced together in collisions and rarely injected independently.

	To compare these results to the Laughlin state, we calculate the two-photon populations $\rho_{jk}$ with angular momenta $j$ and $k$ by extracting the rates of coincidence events, in which two photons are observed nearly simultaneously (Fig.~\ref{fig:AngularMomentumSpace}f; Methods~\ref{methods:AngMomCorr}).
	Coincidence events corresponding to $\ket{66}$ and $\ket{39}$ account for $85(15)\%$ of all observed photon pairs, consistent with angular momentum conservation. Moreover, the ratio $\rho_{39}/\rho_{66}=1.5(5)$ of the pair populations is compatible with the expected ratio of $2.1$ for the Laughlin state. Note that, as a result of our Floquet scheme, polaritons with angular momentum $l=3~\&~9$ are more photon-like than those with $l=6$, such that the observed photonic pair populations are related but not equivalent to the polaritonic pair populations, as detailed in SI~\ref{SI:PhotonsVsPolaritons}; in this work we are assembling \emph{photonic} Laughlin states, not \emph{polaritonic} ones.

	We next test for the remaining essential physical feature of Laughlin states: the particles should avoid each other in real space. This spatial anticorrelation is what minimizes the interaction energy, causing Laughlin-like states to arise in fractional quantum Hall systems; however, it has not been directly observed in existing (electronic) systems.

	To measure spatial correlations we filter the photons exiting the cavity with a single-mode optical fiber (Fig.~\ref{fig:SpatialCorrelation}a). This fiber only admits photons at the location of its tip. Thus, to count photons at a particular position, we simply translate the fiber tip to that position. Since the average density in a state composed of $\ket{39}$ and $\ket{66}$ forms a smooth annulus, we translate the fiber to the radius with the highest density (Fig.~\ref{fig:SpatialCorrelation}b). A natural method for measuring angular correlations $g^{(2)}(\phi, \tau=0)$ between photons separated by the angle $\phi$ around the annulus would be to use two fibers at different positions; our Floquet scheme enables an equivalent measurement using a single fiber, whilst inducing a mode-dependent phase shift between polaritons and photons that we compensate using a linear optical transformation before the fiber (Methods~\ref{methods:SpaceCorr}).

	After phase compensation, two photons that simultaneously exit the cavity rarely appear at the same position (Fig.~\ref{fig:SpatialCorrelation}c). The spatial correlations oscillate with the angle $\phi$ between the photons, with a periodicity of $120^\circ$ arising because our Landau level only includes every third angular momentum state. The observed correlations are compatible with the expected form of this Laughlin state, $g^{(2)}_{L}(\phi)\propto|\psi_L(z_2~=~e^{i\phi}z_1)|^2\propto\sin^4(3\phi/2)$ (Methods~\ref{methods:SpaceCorr}). Thus, we see that while the average density exhibits no angular structure, photon pairs are extremely likely to be separated by about $60^\circ$ (equivalently, $180^\circ$ or $300^\circ$), and very unlikely to be in the same location.

	In addition to directly revealing an important physical property of Laughlin states, the observed spatial correlations are essential for tomographically confirming that the photon pairs have formed Laughlin states. 
	While the detected pair-populations in angular momentum space are suggestive of Laughlin physics, they are insensitive to the phase, or even the purity, of the superposition between $\ket{39}$ and $\ket{66}$. On the other hand, the observed spatial anticorrelation only occurs for a coherent superposition with a minus sign, verifying the formation of a Laughlin state. Taken together, these data indicate that the photon pairs have $76(18)\%$ overlap with the pure Laughlin state, limited primarily by our conservative assumptions about the unmeasured momentum-non-conserving pair populations (Methods~\ref{methods:DensityMatrix}).

We have produced the paradigmatic fractional quantum Hall ground state, the Laughlin state, and measured for the first time the constitutive quantum correlations in two conjugate bases. This work establishes quantum many-body optics as a critical route to breakthroughs in strongly-correlated materials, enabled by the unique microscopic control of our photonic platform in which system parameters are highly controllable, particles can be injected on-demand with particular spatial modes and energies, and intricate correlations can be characterized in almost any measurement basis using straightforward linear optics. Realizing novel state-preparation schemes~\cite{Grusdt2014, Ivanov2018} such as dissipative stabilization~\cite{Kapit2014, Hafezi2015, Umucalilar2017, biella2017phase, Ma2019} will enable the formation of larger topologically ordered states and thus the exploration of fascinating physics including direct-measurement of the statistical phases of anyons~\cite{paredes2001half, umucalilar2013many,Grusdt2016, Dutta2018} or even non-Abelian braiding in the Moore-Read state~\cite{Regnault2004}.

	\droptocpage
	
	\section{Acknowledgements}
	We thank L. Feng and M. Jaffe for feedback on the manuscript. This work was supported by AFOSR grant FA9550-18-1-0317 and AFOSR MURI grant FA9550-16-1-0323. N.S. acknowledges support from the University of Chicago Grainger graduate fellowship and C.B. acknowledges support from the NSF GRFP.
	
	\section{Author Contributions}
    The experiment was designed and built by all authors. N.S. built the primary cavity. L.C., N.S. and C.B. collected the data. L.C. and N.S. analyzed the data. L.C. wrote, and all authors contributed to, the manuscript. 
	
	\section{Author Information}
	The authors declare no competing financial interests. Correspondence and requests for materials should be addressed to J.S. (simonjon@uchicago.edu). 
	
	\section{Data Availability}
	The experimental data presented in this manuscript is available from the corresponding authors upon request.

\renewcommand{\appendixname}{Methods}
	
\setcounter{equation}{0}
\setcounter{figure}{0}
\renewcommand{\theequation}{M\arabic{equation}}
\renewcommand{\thefigure}{M\arabic{figure}}

\clearpage

\setcounter{secnumdepth}{2}

\section*{Methods}

    \subsection{Making polaritons with cavity Rydberg electromagnetically-induced transparency}
    \label{methods:PolaritonBasics}
    Coupling atomic gases with multiple modes of optical resonators provides exciting opportunities for studying many-body physics \cite{gopalakrishnan2009emergent,wickenbrock2013collective,ritsch2013cold,douglas2015quantum,Leonard2017,Vaidya2018}. Here, we use a multimode optical cavity to generate a synthetic gauge field for light \cite{Ozawa2019,hafezi2013imaging,rechtsman2013photonic,Schine2016, Lim2017,Schine2018}. Making photons interact strongly in this cavity enables us to study strongly correlated fractional quantum Hall states of light \cite{hartmann2006strongly,greentree2006quantum,angelakis2007photon, Cho2008, nunnenkamp2011synthetic, Hayward2012, Hafezi2013,Umucalilar2014}.
    
\begin{figure}
		\centering
		\includegraphics{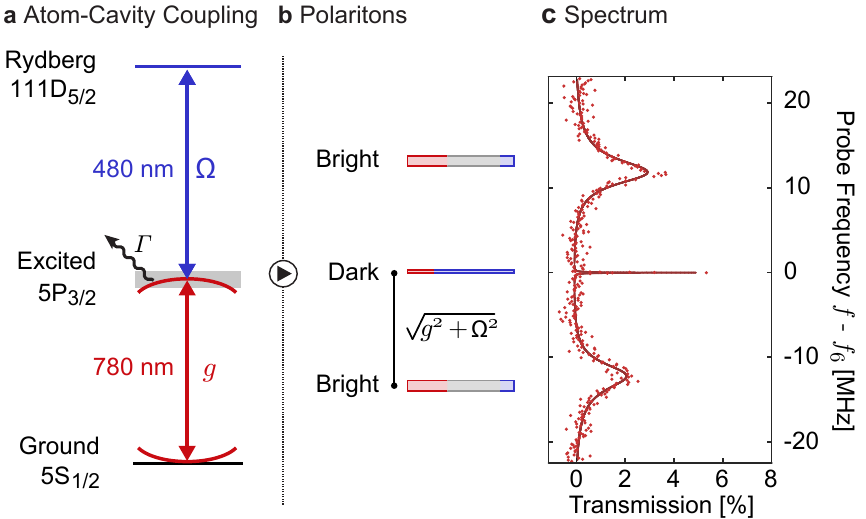} 
		\caption{
			\textbf{a,} Photons at the waist of our optical cavity (red) couple with strength $g$ to the $5S_{1/2}\rightarrow5P_{3/2}$ transition of a gas of ultracold Rubidium atoms, which is subsequently coupled with strength $\Omega$ to the highly excited $111D_{5/2}$ Rydberg state using an additional laser (blue). \textbf{b,}These couplings cause excitations of the atom-cavity system to propagate as polaritons, quasiparticles combining photons with collective atomic excitations.
        	\textbf{c,} The transmission spectrum of the cavity with atoms present directly reveals the narrow dark polariton flanked by two broad bright polariton peaks.
			\label{fig:PolaritonsBasic}
		}
	\end{figure}

    In our system photonic interactions are mediated by Rydberg atoms through a scheme called cavity Rydberg electromagnetically induced transparency~\cite{Fleischhauer2000,Fleischhauer2005,mohapatra2007coherent, pritchard2010cooperative, Guerlin2010,gorshkov2011photon, Peyronel2012,Dudin2012b,Tiarks2014,Gorniaczyk2014,boddeda2016rydberg,Jia2016,Jia2018b}. First, we couple cavity photons with the $5S_{1/2}\rightarrow5P_{3/2}$ transition of the gas of Rubidium atoms at the waist of our primary cavity, which we name the ``science'' cavity, with coupling strength $g$ (Fig.~\ref{fig:PolaritonsBasic}a). The $5P_{3/2}$ state is subsequently coupled to a highly excited Rydberg level ($111D_{5/2}$) by an additional $480$~nm field with coupling strength $\Omega$. As detailed below and in SI~\ref{SI:Cavity}, we use a buildup cavity crossed with the science cavity in order to achieve sufficient intensity of that Rydberg coupling field whilst it covers a large area.
    
    As a result of these couplings, photons no longer propagate in the cavity on their own as they would in vacuum. Instead each cavity mode hosts three kinds of polaritons -- quasiparticles composed of hybrids between a photon and a collective excitation of the atomic gas~\cite{Jia2016} (Fig.~\ref{fig:PolaritonsBasic}b). The nature of these collective excitations is detailed in SI~\ref{SI:CollectiveExcitations}. The two ``bright'' polaritons are largely composed of a photon and a collective $5P_{3/2}$ excitation; we do not employ bright polaritons in this work because they are short lived due to rapid decay of the $5P_{3/2}$ state at $\Gamma=2\pi\times6$~MHz. 
	We utilize dark polaritons throughout the main text of this work. Dark polaritons are purely a superposition of a cavity photon and a collective Rydberg excitation. Because they do not have any $5P_{3/2}$ component, they are long-lived. Moreover, dark polaritons inherit strong interactions from their large collective Rydberg component~\cite{Saffman2010, Jia2018b}. 
    
    A typical scan of the polariton features corresponding to a single cavity mode is shown in Fig.~\ref{fig:PolaritonsBasic}c. That spectrum, taken of the $l=6$ cavity mode, directly reveals the narrow dark polariton excitation at frequency $f_6$ with a linewidth of $60$~kHz, flanked by the two bright polariton resonances which are $\pm\sqrt{g^2+\Omega^2}$ separated in frequency and have linewidths of about 3.7~MHz. For the displayed data, the coupling strengths are $g=2\pi\times13$~MHz and $\Omega=2\pi\times1.3$~MHz and the effective Rydberg state linewidth (including decoherence effects) is observed to be $\Gamma_R=2\pi\times50$~kHz. 

   			\subsection{Experiment Setup}
	\label{methods:ExperimentSetup}
The primary cavity used in our experiments, which couples with the $5S_{1/2}\rightarrow5P_{3/2}$ transition, is the so-called ``science'' cavity, which consists of four mirrors in a twisted (non-planar) configuration~\cite{Schine2016}. The cavity finesse is $\mathcal{F}=1950$ yielding a linewidth of $\kappa=2\pi\times1.4$~MHz.  The modes are approximately Laguerre-Gaussian in the lower waist where they intersect with the atomic cloud; the fundamental mode has a lower waist size of $19$~$\mu$m. These parameters yield a peak coupling strength with a single atom of $g_\mathrm{single}=2\pi\times0.58$~MHz, corresponding to a cooperativity of $\eta=0.16$ per atom~\cite{tanji2011interaction}. The science cavity is crossed with a buildup cavity for increasing the intensity of the Rydberg coupling beam; the buildup makes it possible to achieve a sufficient intensity over the wide area spanning the science cavity modes up to $l=9$ in which we create polaritons. Note that the polarization eigenmodes of both cavities are circular. 
	For more details on the cavity structure, see SI~\ref{SI:Cavity}. 
	
	The science cavity was designed to be tuned to a length at which every third angular momentum state would be degenerate, forming a photonic Landau level~\cite{Schine2016}. However, as detailed in SI~\ref{SI:Cavity}, we found that intracavity aberrations destabilized the cavity modes when the length was tuned to this degeneracy point. Therefore, we instead tuned the cavity length far enough from degeneracy to create a $70$~MHz splitting between every third angular momentum state and utilized the Floquet scheme described below in Methods~\ref{methods:FloquetScheme} to create polaritons which are protected from the effects of intracavity aberrations (SI~\ref{SI:FloquetProtectsAberrations}). 
	
	We mediate interactions between the photons in the science cavity modes using a gas of 6000(1000) cold Rubidium atoms loaded at the lower waist of the cavity; this atom number covers the first ten modes of the science cavity. The gas is cooled to a temperature below $1~\mu$K and polarized into the lowest energy spin-state $\ket{F=2, m_F=-2}$ within the hyperfine manifold $F$ using degenerate Raman sideband cooling~\cite{Kerman2000}. For more details on the atom trapping configuration, atomic polarization, and experiment procedure, see SI~\ref{SI:ExperimentSetup}.

	\begin{figure}
		\centering
		\includegraphics[width=0.8\columnwidth]{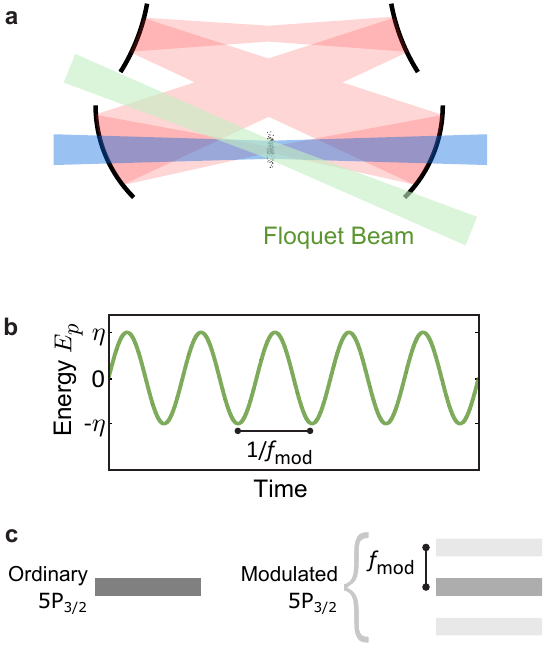} 
		\caption{
			\textbf{Essential features of the Floquet scheme.} 
			\textbf{a,} Our Floquet scheme utilizes an additional laser beam (green) incident upon the atoms with a wavelength of $\lambda=1529$~nm close to the $5P_{3/2}\rightarrow4D$ transition. \textbf{b,} This beam induces a sinusoidally modulated AC Stark shift $E_p=\eta\sin(2\pi f_\mathrm{mod}t)$ of the $5P_{3/2}$ state with amplitude $\eta$ and frequency $f_\mathrm{mod}$. \textbf{c,} As a result of this modulation, the ordinary $5P_{3/2}$ state is split into three bands with energies separated by the modulation frequency. The additional bands enable the atoms to couple with cavity photons at frequencies shifted by $\pm f_\mathrm{mod}$ from the ordinary $5S_{1/2}\rightarrow5P_{3/2}$ resonance frequency.
			For more details on the Floquet scheme see Ref.~\cite{Clark2019}.
			\label{fig:FloquetScheme}
		}
	\end{figure}
	
	\begin{figure*}
		\centering
		\includegraphics{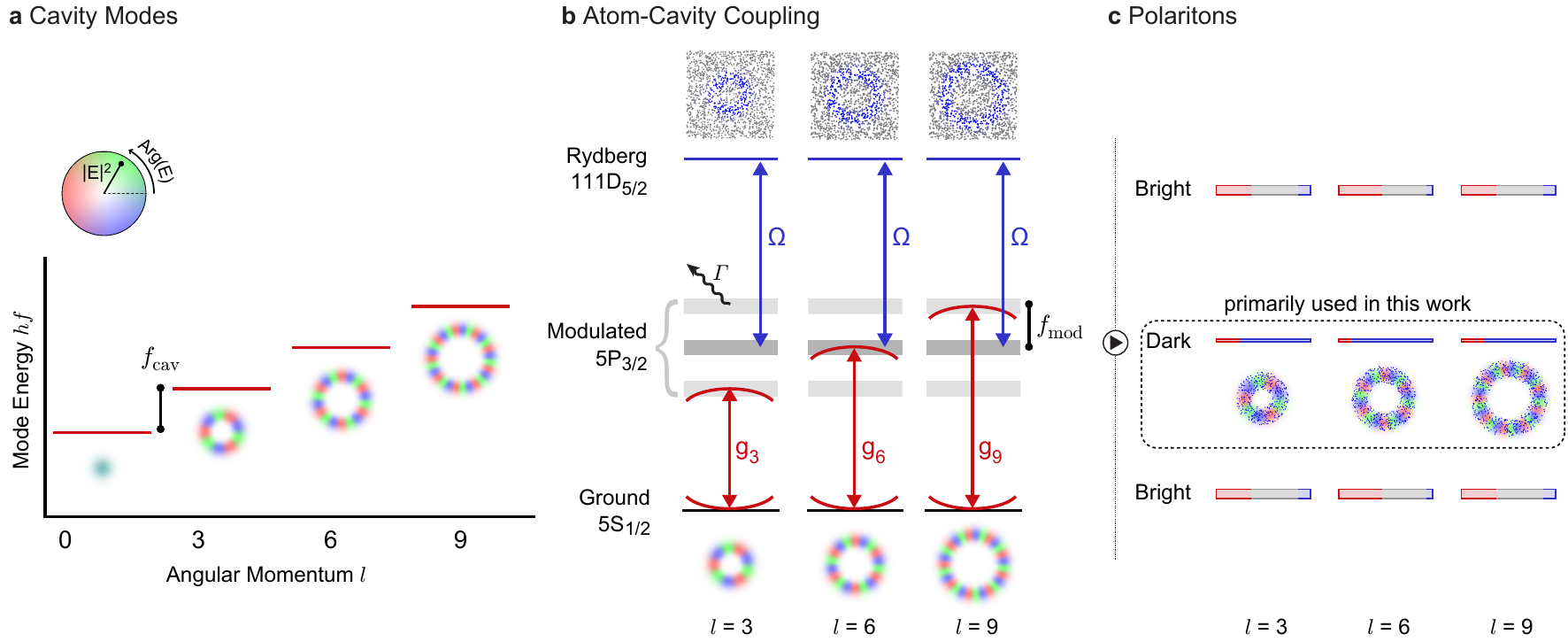} 
		\caption{
			\textbf{Scheme for forming Landau level of Floquet polaritons.} 
			\textbf{a,} The bare cavity modes are not degenerate in this work, but instead the length of the cavity is increased so that there is a $f_\mathrm{cav}\approx70$~MHz splitting between every third angular momentum mode. \textbf{b,} In order to form polaritons in three modes, even though only the $l=6$ mode is resonant with the un-modulated $5S_{1/2}\rightarrow5P_{3/2}$ transition, we utilize the Floquet scheme depicted in Fig.~\ref{fig:FloquetScheme}~\cite{Clark2019}. Modulating the $5P_{3/2}$ state at $f_\mathrm{mod}\approx70$~MHz splits it into three bands (gray), each of which is resonant with one of the three chosen cavity modes. The coupling strengths $g_l$ to each mode $l$ are controlled by the modulation amplitude; in this work, $g_3=g_9=0.37(4)g_6$. Note that each mode couples to a unique collective atomic excitation, as depicted at the top (blue atoms are included in the corresponding collective excitation, while gray atoms are not). \text{c,} This scheme produces polaritons in the $l~=~3,~6,~\&~9$ modes. The dark polaritons can be made effectively degenerate (see Fig.~\ref{fig:EITSpectraFlattening}) without making the corresponding cavity modes degenerate, which protects the polaritons from intracavity aberrations (see SI~\ref{SI:FloquetProtectsAberrations}).
			\label{fig:FloquetPolaritonsFull}
		}
	\end{figure*}

	\begin{figure*}
		\centering
		\includegraphics{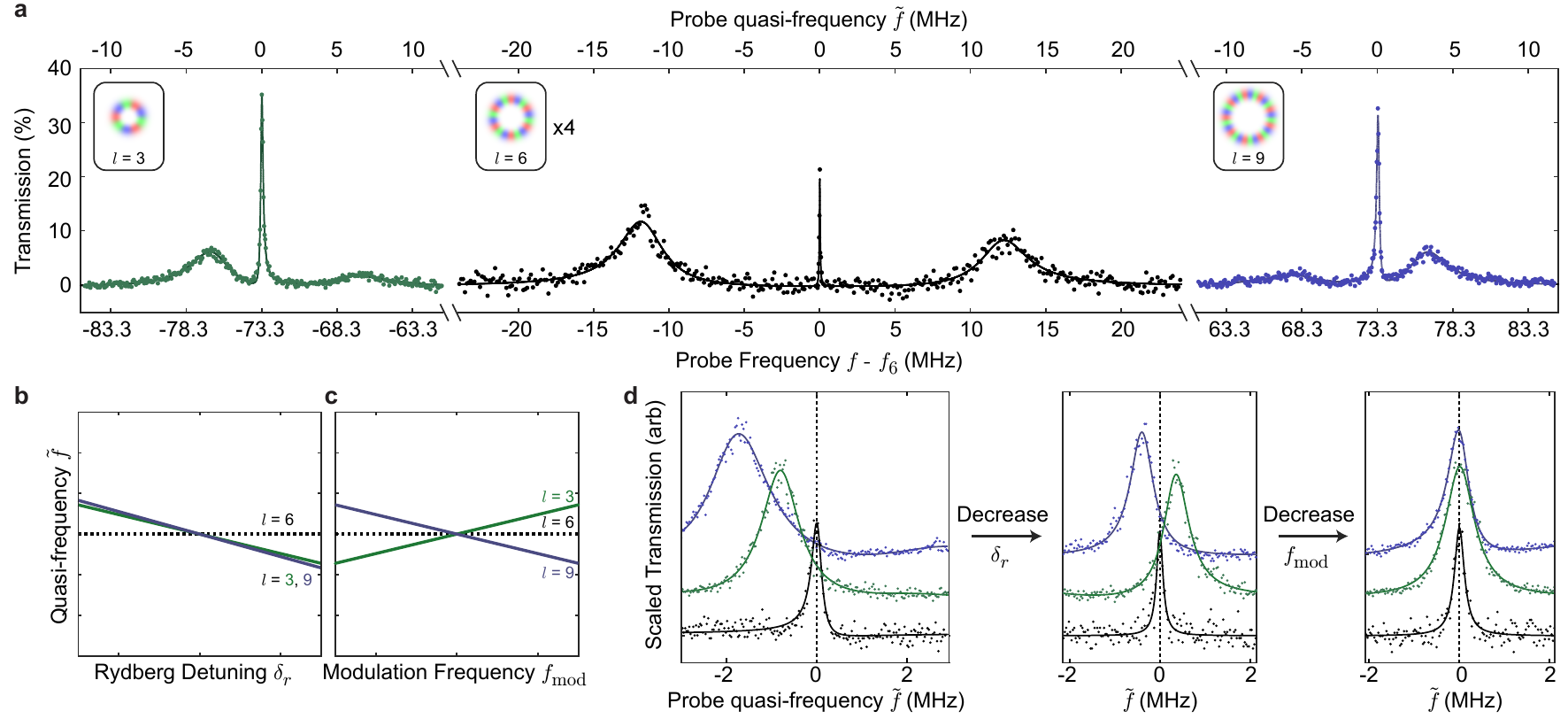} 
		\caption{
			\textbf{Understanding and controlling polariton spectra with the Floquet scheme.} 
			\textbf{a,} Cavity transmission spectrum in the presence of the modulated atoms (see Fig.~\ref{fig:FloquetPolaritonsFull}). The spectrum was collected in three parts, corresponding to injection of photons into $l=3$ (left, green), $l=6$ (middle, black), and $l=9$ (right, violet). Dark polaritons in $l=6$ are less photon-like than those in the other two modes, reducing their relative transmission (SI~\ref{SI:PhotonsVsPolaritons}); we multiply the $l=6$ transmission by four to improve visibility. The lower x-axis indicates the frequency $f$ of the probe laser relative to the $l=6$ dark polariton resonance at $f_6$. The top x-axis indicates the quasi-frequency $\tilde{f}$, proportional to the quasi-energy of the polaritons from a treatment using Floquet theory; $\tilde{f}$ is equal to $f$ modulo the modulation frequency $f_\mathrm{mod}$. \textbf{b}~\&~\textbf{c}, Illustration of the theoretical dependence of the quasi-frequencies of the three dark polariton features on the Rydberg beam detuning (\textbf{b}) and the modulation frequency $f_\mathrm{mod}$ (\textbf{c}). \textbf{d,} Example transmission spectra for the $l=6$ (black, lower), $l=3$ (green, middle), and $l=9$ (violet, upper) dark polaritons as a function of quasi-frequency. The scans are scaled to make their heights equal and have additional vertical offsets for clarity. Shortly before performing each of the experiments reported in the main text, we collect a sequence of plots similar to those displayed here and adjust the Rydberg detuning and modulation frequency to make all three dark polaritons have the same quasi-frequency (right-most plot). The only experiments reported in the main text which did not use this sequence are those shown in Fig.~2b, where instead we varied $\delta_r$ to intentionally vary the energy mismatch between the polaritons. Throughout this figure quasi-frequencies $\tilde{f}$ are reported relative to the $l=6$ dark polariton resonance $\tilde{f}_6$.
			\label{fig:EITSpectraFlattening}
		}
	\end{figure*}
	
	\subsection{Forming the Landau level with Floquet polaritons}
	\label{methods:FloquetScheme}
	This section explains our scheme for forming the Landau level with Floquet polaritons, which provides a number of crucial features that made this work possible. 
	Most importantly, Floquet polaritons helped us mitigate the effects of intracavity aberrations. As detailed in SI~\ref{SI:Cavity}, aberrations prevented us from making the bare cavity modes with angular momenta $l=3~,~6,~\&~9$ degenerate. AT the cavity length which would otherwise make them degenerate, aberrations mix the modes together, splitting them apart in energy and causing them to decay rapidly. Thankfully, Floquet polaritons are protected from the broadening caused by intracavity aberrations, as detailed in SI~\ref{SI:FloquetProtectsAberrations}. The Floquet scheme also provides the ability to efficiently measure angular correlation functions as detailed in Methods~\ref{methods:SpaceCorr} and enables frequency discrimination to improve the selectivity of our angular momentum sorters discussed in Methods~\ref{methods:ModeSorting}. 
	
	The essence of our Floquet scheme is depicted in Fig.~\ref{fig:FloquetScheme} and detailed in Ref.~\cite{Clark2019}. Briefly, a sinusoidal modulation of the energy of the intermediate $5P_{3/2}$ state splits it into three bands which exist at energies separated by the modulation frequency $f_\mathrm{mod}$. Cavity photons can excite the atom through any of these bands, thereby enabling the atoms to couple with modes whose energies are also split by $f_\mathrm{mod}$. 

	To implement the Floquet scheme, we first used temperature and piezo tuning to increase the length of the science cavity and induce a splitting of $f_\mathrm{cav}\approx~70$~MHz between every third angular momentum mode (Fig.~\ref{fig:FloquetPolaritonsFull}a). Then, to enable the $l=3~,~6,~\&~9$ modes to simultaneously support dark polaritons, we fine-tuned the cavity length to make the $l=6$ mode degenerate with the bare $5S_{1/2}\rightarrow5P_{3/2}$ transition and modulated the intermediate state at $f_\mathrm{mod}\approx~70$~MHz to create sidebands resonant with the $l=3~\&~9$ states (Fig.~\ref{fig:FloquetPolaritonsFull}b). By making it possible for the atoms to strongly couple with each of the cavity modes, the Floquet scheme enables all three modes to host polaritons (Fig.~\ref{fig:FloquetPolaritonsFull}c). Note that the splitting $f_\mathrm{cav}$, and therefore the required modulation frequency $f_\mathrm{mod}$, varied over a couple of MHz from day to day due to variation in the cavity temperature. 
	
	Cavity spectroscopy reveals the polariton eigenmodes in all three modes simultaneously (Fig.~\ref{fig:EITSpectraFlattening}a). Because the cavity modes are separated by $f_\mathrm{cav}\approx 70$~MHz, the photonic components of the polaritons also remain separated by approximately $f_\mathrm{cav}$. Therefore, to spectroscopically characterize the set of polaritons for mode $l$, we scan the probe frequency near the frequency $f_l$ of that mode and simultaneously use a digital micromirror device to spatially mode match the probe laser with the cavity mode~\cite{zupancic2016ultra}, producing the spectra shown in Fig.~\ref{fig:EITSpectraFlattening}a. The asymmetric heights and frequency splittings of the bright polaritons relative to the dark polariton for $l=3~\&~9$ are caused by shifts due to coupling with off-resonant bands of the $5P_{3/2}$ state. Moreover, the smaller splitting of the bright polaritons with $l=3~\&~9$ (the ``sideband'' modes) relative to $l=6$ (the ``carrier'' mode) results from smaller atom-cavity couplings $g_{3}\approx g_9=0.35~g_6$ on those modes relative to the central mode. The relative strengths of these couplings are determined by the modulation amplitude; here, we chose to make the sideband couplings weaker in order to make the corresponding dark polaritons more photon-like, see SI~\ref{SI:PhotonsVsPolaritons} for details. 
	
	While the spectroscopic features are all clearly separated in the scan of the probe frequency $f$, the dark polaritons can be tuned to have equal quasi-frequency $\tilde{f}$, where $\tilde{f}=(f \mod f_\mathrm{mod})$ is the probe frequency modulo the modulation frequency. In a Floquet model the energy is only defined up to multiples of the modulation energy; thus, two states with the same quasi-energy (equivalently, quasi-frequency) behave as if they are degenerate~\cite{Eckardt2017}. In the example shown in Fig.~\ref{fig:EITSpectraFlattening}a, the dark polaritons are separated in probe frequency by an amount equal to the modulation frequency $f_\mathrm{mod}=73.3$~MHz; thus, their quasi-frequencies are identical.  
	
	It is not trivial to find conditions at which the quasi-frequencies of the dark polaritons are equal. The primary challenges are the anharmonicity in the cavity mode spectrum and the off-resonant shifts caused by weak couplings to non-resonant Floquet bands. These effects typically prevent the dark polariton energies from matching under conditions which might naively seem suitable; in particular, when $f_\mathrm{cav}=f_\mathrm{mod}$, the quasi-frequency $\tilde{f}_9$ ($\tilde{f}_3$) of dark polaritons with $l=9$ ($l=3$) will typically be too large (small) due to the off-resonant shifts~\cite{Clark2019}. 
	
	The dark polariton quasi-frequency $\tilde{f}_l=\tilde{\delta}_c^l\cos^2{\theta_l}+\tilde{\delta}_r\sin^2{\theta_l}$ in each mode is a weighted average of the cavity mode quasi-frequency $\tilde{\delta}_c^l$ with the Rydberg quasi-frequency $\tilde{\delta}_r$; the weight is determined by the dark-state rotation angle $\theta_l$, which satisfies $\tan{\theta_l}\equiv g_l/\Omega$ (see SI~B5~in~Ref.~\cite{Clark2019}). A smaller ratio $g_l/\Omega$ increases the contribution from the cavity photon and thus makes the polariton more ``photon-like''; in the opposite case, the polariton is more ``Rydberg-like''~\cite{Jia2016}. 
	
	We adjust the Rydberg quasi-frequency $\tilde{\delta}_r$ and the modulation frequency $f_\mathrm{mod}$ in order to tune the dark polaritons into degeneracy. The Rydberg quasi-frequency is controlled by the frequency of the Rydberg coupling laser. Because the $l=6$ polaritons are more Rydberg-like than the $l=3~\&~9$ polaritons, their quasi-frequency increases more rapidly with $\tilde{\delta}_r$, leading to the dependence illustrated in Fig.~\ref{fig:EITSpectraFlattening}b. Note that, in Fig.~2b of the main text, we varied the Rydberg quasi-frequency $\tilde{\delta}_r$ in order to scan the energy mismatch. Adjusting the modulation frequency $f_\mathrm{mod}$ shifts the cavity mode quasi-frequencies $\tilde{\delta}_c^3$ and $\tilde{\delta}_c^9$ oppositely, because they couple to the atoms through opposite modulation sidebands of the $5P_{3/2}$ state (Fig.~\ref{fig:FloquetPolaritonsFull}c).
	
	These two adjustment parameters are sufficient to tune the three dark polariton features into degeneracy, as demonstrated in Fig.~\ref{fig:EITSpectraFlattening}d. In practice, to reach degeneracy we repeatedly measure the quasi-frequency spectrum of the dark polaritons while adjusting parameters. We modify the frequency of the Rydberg laser until the average of the sideband quasi-frequencies matches the carrier, i.e. $(\tilde{f}_3+\tilde{f}_6)/2=\tilde{f}_6$. Similarly, we vary $f_\mathrm{mod}$ until $\tilde{f}_3=\tilde{f}_9$. In the end, we are able to satisfy these conditions to an accuracy well within the linewidths of the dark polaritons.

	\begin{figure}
		\centering
		\includegraphics[width=0.8\columnwidth]{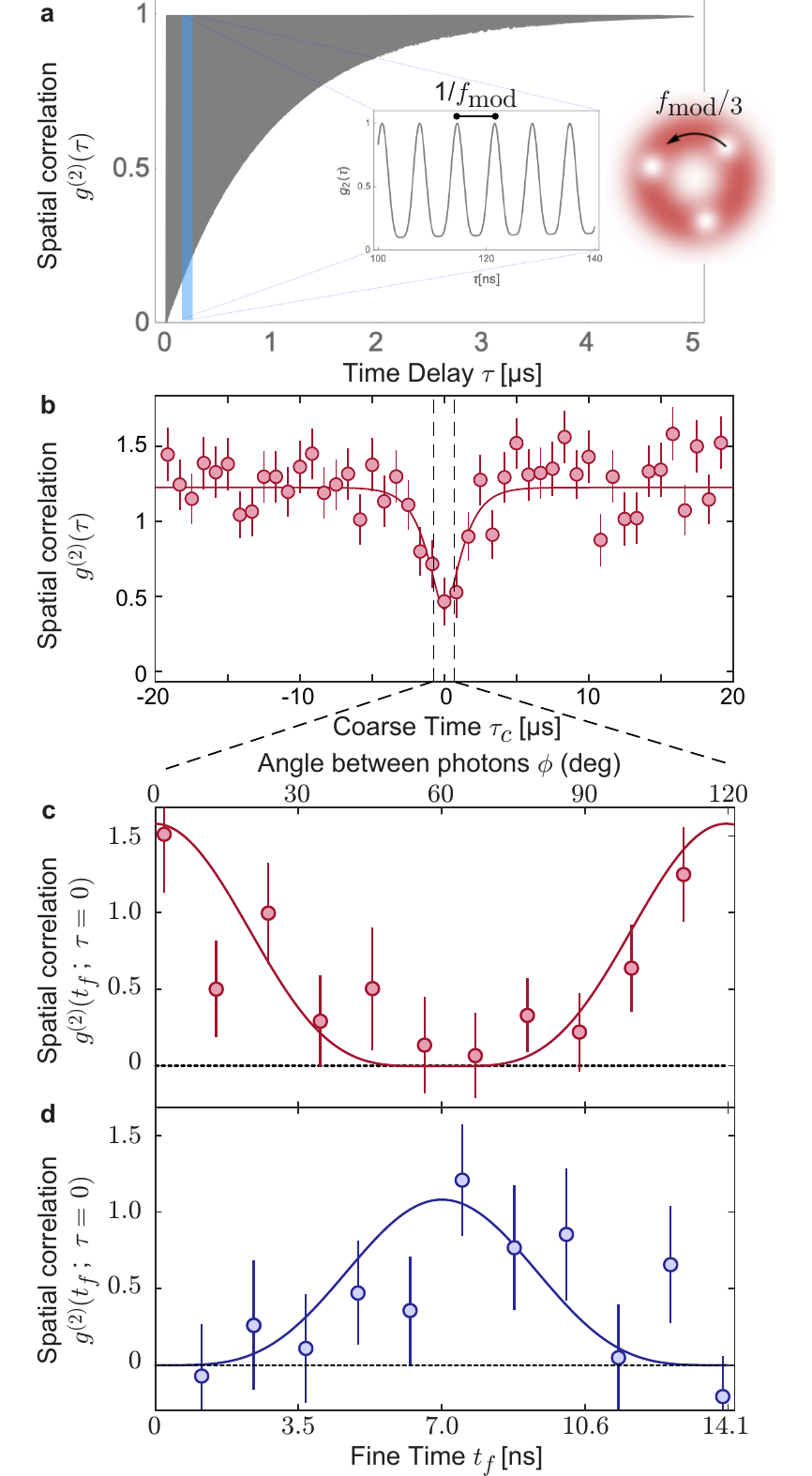} 
		\caption{
			\textbf{Extracting angular correlations from rapid rotation of the second photon.} 
			\textbf{a,} Illustration of the correlation function measured through a single mode optical fiber. Rapid oscillation in the signal at $f_\mathrm{mod}$ arises from the wavefunction of the second photon, which has three holes whose positions are determined by that of the first detected photon and which rotate at $f_\mathrm{mod}/3$ (inset). The rotation signal decays over a microsecond timescale containing hundreds of rotation periods.
    		\textbf{b,} When viewed in coarse time $\tau_c$ with averaging over a timescale much longer than the rotation period, the correlations through the single mode fiber exhibit ordinary antibunching.
    		\textbf{c,} Breaking up the central data point into 11 parts based on their fine time $t_f$ within the rotation period reveals the oscillation of the correlations. The instantaneous correlation ($t_f=0$) corresponds to the same-location correlation ($\phi=0$), while half a period later we see the correlations between locations separated by $\phi=60^\circ$. Without the phase compensator, we observe photon bunching at $t_f=0$. \textbf{d,} With the phase compensation cavity implemented (see Methods~\ref{methods:SpaceCorr}), photons are anticorrelated at $t_f=0$.
			\label{fig:SpatialCorrelationsFineTime}
		}
	\end{figure}

	\begin{figure*}
		\centering
		\includegraphics{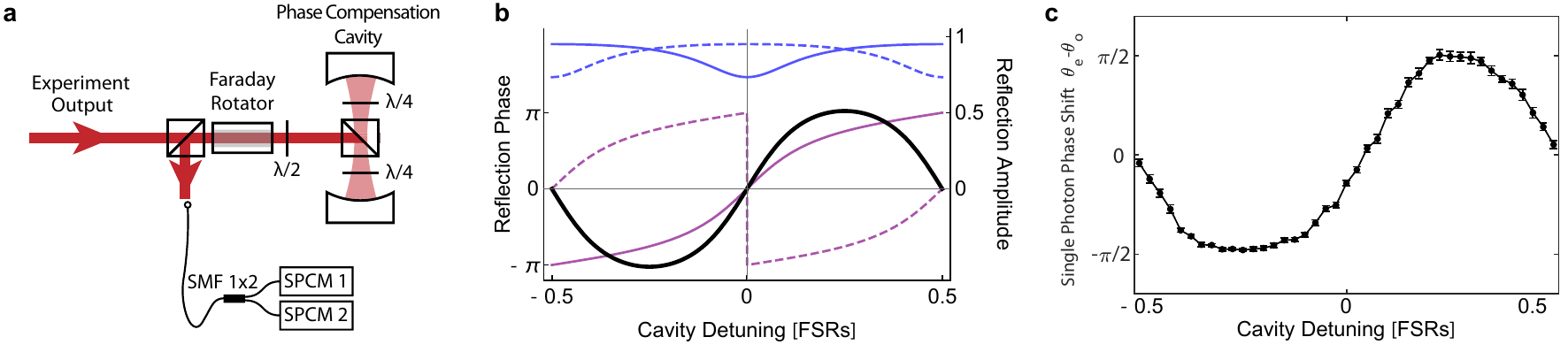} 
		\caption{
			\textbf{Phase compensation cavity for spatial correlation measurements.} 
			\textbf{a,} To compensate for the additional phase shift of the photons relative to the polaritons which is induced by the Floquet scheme, we send the light emitted from the main science cavity (``experiment output'') to a phase compensation cavity (PCC), which contains two quarter waveplates and a polarizing beam splitter cube that can be used to tune its finesse. Almost all of the light is reflected from the PCC after experiencing the desired phase shift. It subsequently travels to the single mode fiber, where it is split and sent to two separate single photon counting modules (SPCM) for correlation measurements.
			\textbf{b,} The PCC is a confocal cavity with 82\% effective transmission of the first mirror (determined by the waveplate angle), 0\% transmission of the second, and 12\% intracavity loss. While all modes experience relatively little loss (blue, top), even modes (solid) reflected from the cavity experience a different phase shift (purple) than odd modes (dashed). The net phase shift for the two-photon states $\ket{66}$ relative to $\ket{39}$ (black solid) can be tuned over $2\pi$ by changing the cavity length.
			\textbf{c,} Experimental calibration in the \textit{single-photon sector} verifies a tunable phase shift of up to $\pm\pi/2$ for even modes relative to odd modes. 
			\label{fig:PhaseShiftingCavity}
		}
	\end{figure*}

	\subsection{Angular momentum mode sorting}
	\label{methods:ModeSorting}
	To perform detection of photons in angular momentum space we utilize the mode sorting setup depicted in Fig.~2a. The key elements in this setup are the mode filters, each consisting of a tunable two mirror non-degenerate cavity. These filter cavities are independently locked to a 795~nm laser with tunable frequency offsets, such that we can match the frequency of the filter cavity mode with angular momentum $l$ to the frequency of the photons exiting the science cavity with the same angular momentum. The three cavities differ slightly in their parameters, but the typical cavity is composed of two identical mirrors with $98.7\%$ transmission and radii of curvature $R=10$~m at a separation of $15$~cm, yielding a linewidth of $\kappa_{filt}=4$~MHz and a transverse mode spacing of 55~MHz. Because the science cavity angular momentum modes are separated by $f_\mathrm{cav}\approx70$~MHz much greater than the filter cavity linewidth, the filters are able to discriminate between the science photons using both their spatial mode structure and their frequency. The typical mode sorting/filter cavity has a net transmission of $40\%$ for photons with the desired angular momentum and suppresses transmission of photons with the wrong angular momentum by a factor of 2000. The transmission of each mode sorting cavity is routed to a unique single photon counting module by a multimode fiber.
	
	The experiments reported in Fig.~2 of the main text were performed using all three mode sorting cavities simultaneously, with each cavity tuned to transmit a different angular momentum mode. The experiments reported in Fig.~3 were performed before the third mode sorting cavity was constructed. Therefore, the data shown in Fig.~3d were collected in separate experiment runs from the data in Fig.~3e; in between, the two mode sorters were adjusted to transmit the relevant angular momentum modes.

	\subsection{Two-Photon Correlations}
	\label{methods:CorrelationsG2}
	We characterize the light exiting the science cavity using two-photon correlation functions,
    \begin{equation}
    g^{(2)}_{jk}(\tau)=\frac{\left< n_j(t)n'_k(t+\tau)\right>_t}{ \left<n_j\right>\left< n'_k\right>},
    \label{eqn:g_definition}
    \end{equation} between the photons $n_j(t)$ in mode $j$ counted at a first detector and $n'_k(t)$ in mode $k$ counted at a second detector, where the angle brackets denote time averaging and $\tau$ is the time delay between the detection events. The mode labels $j$ and $k$ may represent angular momentum modes with $l=j$ and $l=k$, or they may represent the spatially localized mode of the single-mode fiber. When $j=k$, we typically use a beam splitter to divide the photons between two separate single photon counting detectors. For many of our measurements the background count rates in the absence of the probe are comparable to the signal rates in the presence of the probe; the correlation functions presented in this work correspond to the correlations of signal photons with the backgrounds removed, as detailed in SI~A2~of~Ref.~\cite{Clark2019}.

	\subsubsection{Angular Momentum Correlations}
	\label{methods:AngMomCorr}
	The correlation functions presented in the main text Fig.~3c-e were collected during three separate runs of the experiment. In the first run (Fig.~3c), a multi-mode fiber splitter was used to collect all of the photons exiting the science cavity, regardless of mode, and split them evenly between two single photon counters. In the second run (Fig.~3d), the first mode sorting filter was tuned to transmit half of the photons in $l=6$, while the second mode sorting filter transmitted the remaining half of photons with $l=6$. In the third run (Fig.~3e), the first mode sorting filter transmitted photons with $l=9$ and the second filter transmitted photons with $l=3$. Note that the local maximum at $g^{(2)}_{all}(0)$ in Fig.~3c appears because the peak in $g^{(2)}_{39}$ is somewhat narrower than the dip in $g^{(2)}_{66}$, as polaritons with $l=3~\&~9$ have shorter lifetimes than polaritons with $l=6$ due to our Floquet scheme (SI~\ref{SI:PhotonsVsPolaritons}). 
    	
    The two-photon population fractions shown in Fig.~3f of the main text for $j=k=6$ and $j=3$, $k=9$ are calculated by comparing the observed coincidence rates as,
    \begin{equation}
        \frac{\rho_{ij}}{\rho_{tot}}=\frac{\left< n_j(t)n'_k(t)\right>_t}{\left< n_{mm}(t)n'_{mm}(t)\right>_t}\frac{2}{(1+\delta_{jk})\xi_j\xi'_k},
    \end{equation}
    where $n_{mm}$ ($n'_{mm}$) is the signal at the first (second) output port of the multimode splitter used to measure $g^{(2)}_{all}(\tau)$, $\xi_j$ ($\xi'_k$) is the mode-sorted detection efficiency for mode $j$ ($k$) on the first (second) detector relative to the efficiency of a single output port of the multimode splitter, and $\delta_{jk}$ is the Kronecker delta. Each of the detection efficiencies is independently calibrated by probing on the relevant dark polariton resonance and comparing the count rate after the mode-sorting filter to the count rate after the multimode splitter. Let $N_{jk}(\tau)=\left< n_j(t)n'_k(t+\tau)\right>_t$ be the observed rate of two-photon events separated by time $\tau$. For the mode-sorted cases we estimate $N_{jk}(0)$ by fitting the observed delay-time-dependent $N_{jk}(\tau)$ and extracting the zero-delay value from the fit; empirically, we find that $N_{39}(\tau)$ is well fit by a lorentzian and $N_{66}(\tau)$ is well fit by the solution to the optical Bloch equations for a two-level system (see Ref.~\cite{Clark2019}~SI~A2). For the multimode case we do not have a suitable fitting function, and therefore we directly use the observed value $\int_{-\Delta/2}^{\Delta/2}d{\tau}N_{mm}(\tau)$ averaged over a time bin with width $\Delta=600$~ns around zero delay.
    
    We find that the observed populations $\rho_{39}$ and $\rho_{66}$ may not fully account for the observed coincidences in the multimode data, with $\frac{\rho_{39}+\rho_{66}}{\rho_{tot}}=0.85(15)$. 
    Since we do not directly measure the correlations in any other combinations of angular momentum modes, Fig.~3f uses the estimate that the remaining population fraction of $0.15(15)$ is split evenly among the remaining mode pairs. The actual distribution of this additional population among the four possible mode pairs does not affect any of our conclusions, including the density matrix and the overlap of the observed state with the Laughlin state as discussed in Methods~\ref{methods:DensityMatrix}. Physically, we note that angular momentum conservation should prevent these other mode pairs from being populated at all; however, we are not able to place any further constraints on their populations based on our measurements. Note also that our use of a weak probe beam, as well as the strong interaction-induced blockade of all three polariton states among these three modes, should make the effect of three-photon states on these results negligible.

	\subsubsection{Spatial Correlations}
	\label{methods:SpaceCorr}
We perform measurements in real space using a single mode fiber. To properly align the fiber to detect photons in the Landau level at the desired position, we first choose an asphere as well as the position and angle of the fiber which maximizes the coupling of $l=0$ photons from the cavity into the fiber. Once that coupling is optimized, we use a four-axis stage (two-dimensional translation, tip, and tilt) to perform a ``magnetic translation'' of the fiber tip \cite{Schine2016}. In particular, we send $l=6$ photons through the cavity and repeatedly adjust the position and angle of the fiber until the fraction of $l=6$ photons entering the fiber is maximized. This process matches the fiber to a magnetically-translated gaussian in our lowest Landau level at a radius near the peak in the average density of photons that are exiting the cavity in pairs (as depicted in Fig.~4b). 

	 Ordinarily, a single fiber would only be able to reveal the correlations at that single location. However, as discussed in SI~\ref{SI:PhotonsVsPolaritons}, immediately after a first photon is detected through the single mode fiber, the Floquet scheme used in this work causes the second photon to rotate one third of the way around ring with a temporal period of $T_{mod}=1/f_\mathrm{mod}=14$~ns which is much faster than other dynamical timescales of the system. Note that any wavefunction in our Landau level has an azimuthal periodicity of $2\pi/3$, and therefore rotating one third of the way around the ring is sufficient to map the wavefunction back on to itself. Therefore, the correlations between photons in the single mode fiber separated by a ``fine time'' $t_f=\tau \mod T_{mod}$ are equivalent to the angular correlations that would be observed between photons at equal time with angular separation $\phi_f=\frac{2\pi}{3}\frac{t_f}{T_{mod}}$,
	 \begin{equation}
	     g^{(2)}(\phi_f; \tau=0)=g^{(2)}_{ff}(t_f). 
	     \label{eqn:g2TimeToSpace}
	 \end{equation}
	 An illustration of this behavior is provided in Fig.~\ref{fig:SpatialCorrelationsFineTime}a.
	 
	 Therefore, to characterize the angular correlations, we perform high resolution measurements of photon arrival times through the single optical fiber. After entering the single-mode fiber, we split the field into two halves and send each to a single photon counting module. We use a home-built FPGA-based photon time tagger with 1.4~ns resolution to characterize the photon arrival times. We convert the temporal correlations $g^{(2)}_{ff}(t_f)$ into the correlation function $g^{(2)}(\phi, \tau=0)$ using Eq.~\ref{eqn:g2TimeToSpace}. Since we are unable to collect sufficient statistics to characterize correlations during a single rotation period, we instead average data corresponding to the same fine time $t_f$ over many rotation periods within a large ``coarse time'' bin of total duration 1.2~$\mu$s. The result of this process is depicted in Fig.~\ref{fig:SpatialCorrelationsFineTime}b~\&~c. To calibrate any additional time offset due to time delay differences in the two detection paths, we perform a separate set of measurements sending rapidly intensity modulated coherent light directly into the fiber splitter; we extract the time delay as the difference in measured arrival times for peaks and troughs of that signal along the two detection paths. That time delay calibration has been accounted for in Fig.~\ref{fig:SpatialCorrelationsFineTime}c~\&~d, such that the correlation signal at $t_f=0$ properly corresponds to photons arriving at the fiber tip at the same time. 
	
	As explained in SI~\ref{SI:PhotonsVsPolaritons}, our Floquet scheme induces a shift of $\pi$ in the relative phase between $\ket{66}$ and $\ket{39}$ in the photonic state relative to the polaritonic state. As a result, a spatial correlation measurement performed directly after the science cavity reveals photon bunching at $t_f=0$ ($\phi=0$), as shown in Fig.~\ref{fig:SpatialCorrelationsFineTime}c. While this phase shift can be compensated with a linear optical transformation, it is not affected by diffraction, single lenses, or single mirrors, and thus must be manipulated a slightly more sophisticated ``optical element''.  
	
	In order to compensate for this phase and produce a photonic Laughlin state, we implement the phase compensation cavity (PCC) depicted in Fig.~\ref{fig:PhaseShiftingCavity}a. The PCC consists of two curved mirrors with an intracavity polarizing beam splitter and quarter waveplates which can be used to tune the effective reflectivity of the mirrors. We adjust the waveplates to make the PCC behave as a single-ended cavity with 82\% transmission on the input coupler, maximum reflection on the output coupler, and 12\% intracavity loss; this corresponds to an overall finesse of 3. Moreover, we adjust the length of the cavity to be near-confocal, so that all even modes are degenerate and all odd modes are degenerate. While nearly all of the light incident upon the PCC is reflected regardless of its parity or frequency, the PCC induces a tunable, parity-dependent phase shift (Fig.~\ref{fig:PhaseShiftingCavity}b). We chose the finesse of the cavity so that the phase shift per photon for even modes relative to odd modes is $\pi/2$ over a wide range of cavity detunings (Fig.~\ref{fig:PhaseShiftingCavity}c), and we then lock the cavity in the center of that range. 
	
	Because $\ket{66}$ contains two even photons and $\ket{39}$ contains two odd photons, reflecting from the PCC induces a total relative phase shift of $\pi$ between them. Therefore, it exactly compensates for the extra phase induced by the Floquet scheme. As a result, when we reflect the light exiting the science cavity off of the PCC before performing the spatial correlation measurement, we observe the antibunching at time $t_f=0$ ($\phi=0$) which is characteristic of the Laughlin state.

	\begin{figure}
	\centering
	\includegraphics{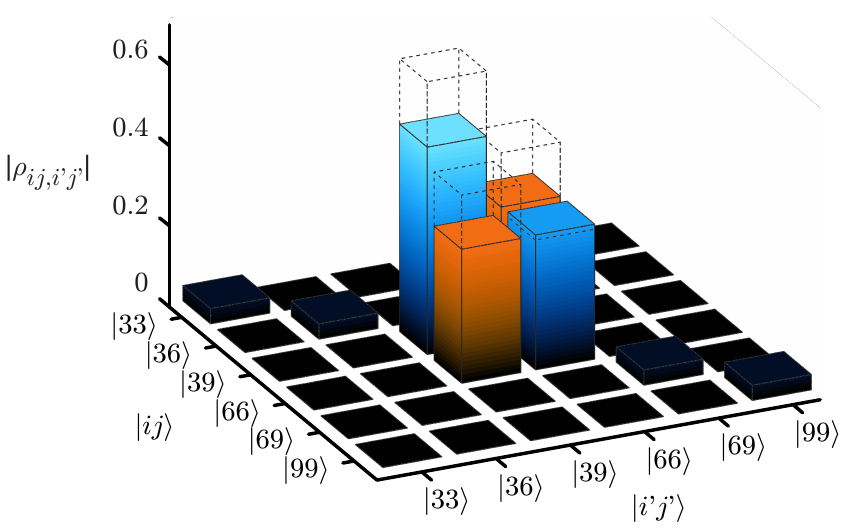} 
	\caption{
		\textbf{Density matrix in the two-photon sector.} 
		Combining our results in real space (Fig.~\ref{fig:SpatialCorrelation}) and angular-momentum space (Fig.~\ref{fig:AngularMomentumSpace}) enables us to reconstruct the density matrix of the two-photon state which exits our system (solid bars). Blue bars are positive, orange bars are negative. Dashed bars correspond to the pure Laughlin state. 
		\label{fig:DensityMatrix}}
	\end{figure}

	\subsection{Density matrix reconstruction}
	\label{methods:DensityMatrix}
	
	Here we combine the results of our measurements in angular momentum space and real space to calculate the density matrix among the two-photon states in our system, which has the elements,
	\begin{equation}
	    \rho_{ij,i'j'}=\bra{ij}\rho\ket{i'j'}.
	\end{equation}
	The calculation of the diagonal elements $\rho_{39}\equiv\rho_{39,39}$ was discussed in Methods~\ref{methods:AngMomCorr} with the results shown in the main text Fig.~3f. 
	
	We determine the relevant off-diagonal density-matrix element $\rho_{39,66}$ based on the observed spatial correlations.
	The angular correlation function of the ideal Laughlin state is,
	\begin{align*}
	    g^{(2)}(\phi)&=\frac{8\bar{g}}{3}\sin^4(3\phi/2)\\
	        &=\bar{g}\left(1+\frac{1}{3}\cos6\phi-\frac{4}{3}\cos3\phi\right),
	\end{align*}
	where $\bar{g}$ is the time-averaged correlation.
	With an arbitrary density matrix among the states $\ket{39}$ and $\ket{66}$, the correlation function takes the form,
	\begin{multline}
	    g^{(2)}(\phi)=\bar{g}\bigg(\frac{651}{260}\rho_{66}+\frac{155}{546}\rho_{39}(1+\cos6\phi)\\
	    -\frac{124}{\sqrt{1890}}\Re(\rho_{39,66})\cos3\phi\bigg).
	    \label{eqn:SpaceCorrFit}
	\end{multline}
	This form is independent of the radius at which the single mode fiber is placed. We fit the spatial correlation data shown in Fig.~\ref{fig:SpatialCorrelationsFineTime}d with Eq.~\ref{eqn:SpaceCorrFit} and use the separately measured populations $\rho_{66}$ and $\rho_{39}$ to extract the coherence $\rho_{39,66}=-0.33(16)$. 
	
	Note that Eq.~\ref{eqn:SpaceCorrFit} neglects the additional possible contributions to the average value which come from the population in other angular momentum pair states ($\ket{33},~\ket{36}$, etc.). Because these contributions depend more sensitively on the position of the fiber, and because we do not know how the additional $15\%$ of the population is distributed among these pair states, we take the most conservative approach and neglect their contributions entirely. This approach is conservative because it yields the smallest magnitude of $\rho_{39,66}$ and therefore the smallest inferred overlap with the Laughlin state. Note also that because only the real part of the coherence $\Re(\rho_{39,66})$ contributes to the observed correlations, we cannot distinguish between a loss in magnitude of $\rho_{39,66}$ and the presence of an imaginary component. We treat our results as if there is no imaginary component; however, this choice does not affect the inferred overlap with the Laughlin state below.  
	
	Our final result for the density matrix based on these conservative assumptions is shown in Fig.~\ref{fig:DensityMatrix}. This density matrix has a ``Laughlin fidelity'', its overlap with the pure Laughlin state, of,
	\begin{equation}
	    F=\bra{L}\rho\ket{L}=76(18)\%,
	\end{equation}
	as reported in the main text.
	
	Alternatively, we can calculate the Laughlin fidelity under an optimistic set of assumptions. In particular, because this system should satisfy angular momentum conservation, we can assume that there is actually no population in the other angular momentum pair states; in that case we set $\rho_{33}=\rho_{36}=\rho_{69}=\rho_{99}=0$, and we re-scale the populations in the remaining two states such that their ratio matches our experimental observations but their sum is $\rho_{66}+\rho_{39}=1$. If we also recalculate the off-diagonal coherence under those assumptions, then the resulting density matrix corresponds to an optimistic Laughlin fidelity of 90(18)\%.

	\renewcommand{\tocname}{Supplementary Information}
\renewcommand{\appendixname}{Supplement}
	
\setcounter{equation}{0}
\setcounter{figure}{0}
\renewcommand{\theequation}{S\arabic{equation}}
\renewcommand{\thefigure}{S\arabic{figure}}

\incltocpage
\clearpage

\tableofcontents
\appendix
\setcounter{secnumdepth}{2}
	
	\begin{figure*}
		\centering
		\includegraphics[width=0.95\textwidth]{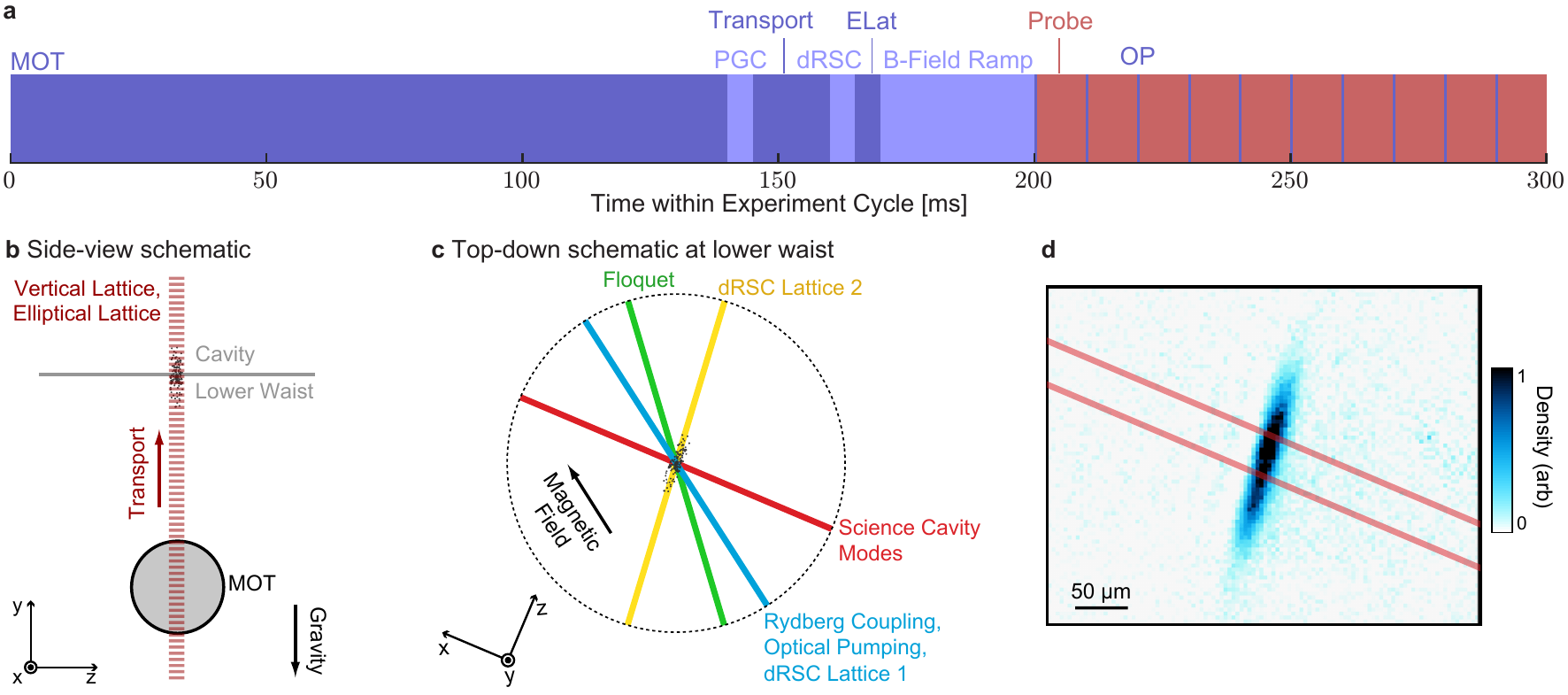} 
		\caption{
			\textbf{Experiment sequence and schematics.} 
			\textbf{a,} Our typical experiment sequence repeats every 300 ms. It begins with loading of the magneto-optical trap (MOT) while the electronic control system prepares to execute the rest of the sequence, followed by polarization-gradient cooling (PGC), transport of the atoms from the MOT to the cavity using a moving, vertical optical lattice (VLat), degenerate Raman sideband cooling (dRSC), transfer of the atoms into an elliptical optical lattice (ELat) to reshape the cloud, a ramp of the bias magnetic field and settling time to ensure stability, and finally ten, $\sim10$~ms probe cycles separated by brief pulses of optical pumping (OP) light. \textbf{b,} A schematic of the experiment shown from the side, indicating beam alignment. \textbf{c,} A schematic of the experiment from a top-down perspective, indicating the four beam paths which cross at the lower waist of the science cavity, as well as the beam(s) which propagate along each path. The bias magnetic field is parallel to the Rydberg coupling axis. \textbf{d,} An example image of the atomic cloud after transfer into the ELat, which is wide transverse to the cavity mode but narrow along the cavity $z$-axis. The red lines indicate the edges of the fundamental mode of the science cavity.
			\label{fig:ExpSequenceAndSchematics}
		}
\end{figure*}

	\section{Experiment}

\subsection{Experiment setup and typical sequence}
\label{SI:ExperimentSetup}

	Our experiments begin with a gas of $10^7$ cold $^{87}$Rb atoms at a temperature of 15~$\mu$K prepared using a magneto-optical trap and polarization gradient cooling (Fig.~\ref{fig:ExpSequenceAndSchematics}a). We transport that gas vertically into the waist of our science cavity using a moving optical lattice (Fig.~\ref{fig:ExpSequenceAndSchematics}b). Previous iterations of the experiment would then release the gas from the trap in order to prevent inhomogeneous broadening of the ground to Rydberg transition due to the trapping beams~\cite{Jia2018b, Clark2019}. To improve the atomic density and experiment duty cycle for these experiments, we now cool the gas to a temperature below $1~\mu$K and polarize the atoms into the lowest energy spin-state $\ket{F=2, m_F=-2}$ within the hyperfine manifold $F$ using degenerate Raman sideband cooling~\cite{Kerman2000}. We then transfer the gas into an elliptical dipole trap, which we retroreflect to form an optical lattice in order to support the atoms against gravity. This trap makes the gas at the cavity waist highly elliptical, with a very thin root-mean-squared (RMS) radius of 11~$\mu$m along the science cavity axis ($\hat{z})$ which makes the polariton interactions strong~\cite{Jia2018b} while maintaining much larger transverse RMS radii of 51~$\mu$m$\times~120~\mu$m in order to cover the area spanned by the desired transverse modes of the cavity (Fig.~\ref{fig:ExpSequenceAndSchematics}d).
	
	The elliptical lattice primarily traps the ground state atoms and only slightly perturbs the Rydberg state, resulting in an AC Stark shift of the two-photon resonance frequency for the $5S_{1/2}\rightarrow111D_{5/2}$ transition of each atom. These shifts lead to inhomogeneous broadening because atoms at different positions in the trap experience different potentials. However, because the sample temperature (and therefore the spread of Stark shifts) of less than $1~\mu$K corresponds to a frequency scale of less than 20~kHz, this inhomogeneous broadening makes a tolerably small contribution to the observed Rydberg linewidth of $\Gamma_R$=$2\pi\times 50$~kHz. Therefore, we are able to leave the elliptical lattice on throughout the probe cycle in which we create dark polaritons.

	We have found that careful management of the polarizations of the atoms and the light is crucial for maximizing the lifetime of our dark polaritons. In particular, while probing the system, we use a $5$~G bias magnetic field directed along the Rydberg coupling beam propagation axis in order to provide a Zeeman splitting within each of the atomic state manifolds (Fig.~\ref{fig:ExpSequenceAndSchematics}c). The sample is also repeatedly optically pumped along that same axis to ensure that all atoms start in the $\ket{F=2, m_F=-2}$ state. Because the Rydberg coupling beam is coaligned with the magnetic field, and the science cavity mode axis is only $\sim30^\circ$ away, the circularly polarized cavity photons only couple the atoms to a single spin state in each manifold. These conditions make our polarized atoms behave much more like a three-level system than an unpolarized sample; the three relevant levels are $\ket{5S_{1/2}, F=2, m_F=-2}$, $\ket{5P_{3/2}, F=3, m_F=-3}$, and $\ket{111D_{5/2}, m_J=5/2}$ where we note that the hyperfine splitting of the Rydberg state is negligible. We have found that this setup maximizes the dark polariton lifetime and minimizes the creation of stray Rydberg atoms. For more details on the experiment setup, see SI~\ref{SI:ExperimentSetup}.
	
	The full experiment procedure is depicted in Fig.~\ref{fig:ExpSequenceAndSchematics}a. Because we are able to trap the atoms while probing, we now spend approximately 200~ms preparing the sample and 100~ms performing science measurements in each trial, for a duty cycle of roughly 33\%; this represents a dramatic improvement over the previous scheme, in which the duty cycle was approximately 1\%~\cite{Jia2018b, Clark2019}. Note that the time spent loading the magneto-optical trap is also used by the computer to prepare the control electronics for running the next iteration of the experiment sequence.

	We have also made a number of technical improvements to the Floquet setup relative to our original implementation in Ref.~\cite{Clark2019}. 
	In this work the laser beam inducing the sinusoidally modulated AC Stark shift of the $5P_{3/2}$ state (the ``Floquet beam'') has a wavelength of $\lambda_f=1529$~nm near the $5P_{3/2}\rightarrow4D_{3/2}$ and $5P_{3/2}\rightarrow4D_{5/2}$ transitions. This new Floquet beam has a much greater detuning from the $5S_{1/2}\rightarrow5P_{3/2}$ at 780~nm than the old Floquet beam at $776$~nm, which was used in our previous work. This increased detuning dramatically reduces the inhomogeneous broadening of the atomic ground state caused by the inhomogeneous intensity of the Floquet beam, thereby maximizing the coherence time of the dark polaritons. We set the frequency components of the multichromatic Floquet field and their relative amplitudes in order to sinusoidally modulate the energy of the $5P_{3/2}$ state while keeping its average energy constant \cite{Clark2019}. In particular, the Floquet laser is locked at a frequency $7$~GHz detuned from the $5P_{3/2}\rightarrow4D_{5/2}$ transition, such that it has nearly equal and opposite detunings from $4D_{5/2}$ and $4D_{3/2}$ which are split by 13.5~GHz. After amplifying the beam with an erbium-doped fiber amplifier we use a fiber electro-optic modulator (EOM) to simultaneously phase-modulate the light at frequencies of $8.500$~GHz and $8.573$~GHz; each of the resulting first order EOM-induced sidebands has a power approximately 1/6 that of the optical carrier. We fine-tune the strength of the sidebands in order to achieve a large modulation amplitude of the $5P_{3/2}$ energy at $f_\mathrm{mod}=73$~MHz while its average energy is unchanged.

\begin{figure}
	\centering
	\includegraphics[width=\columnwidth]{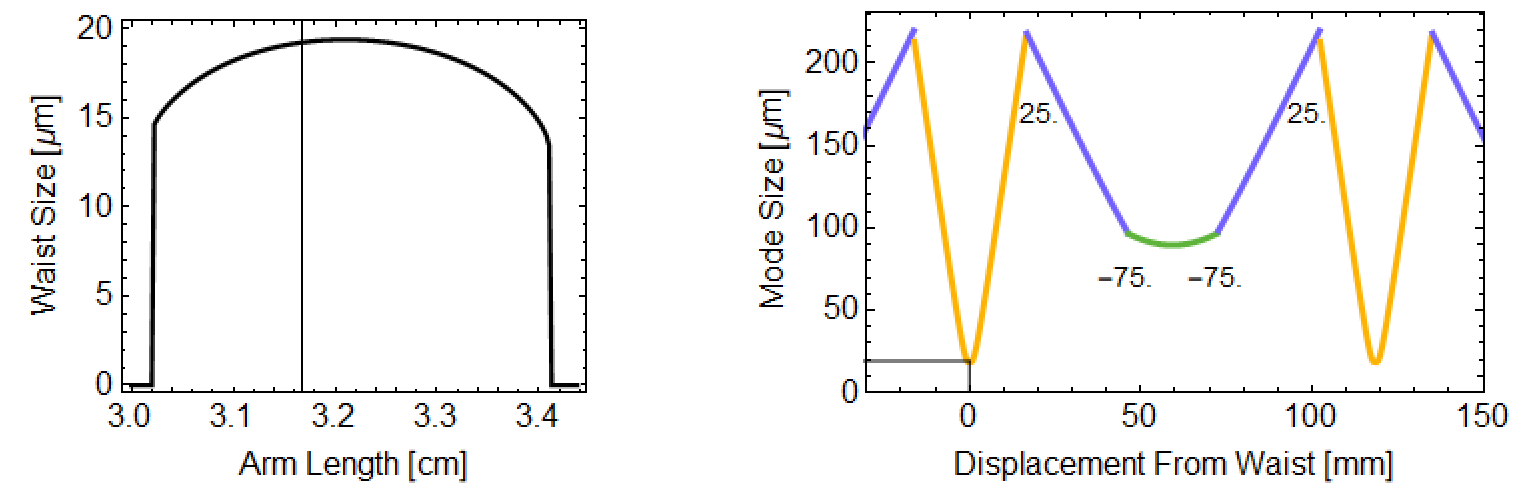}
	\caption{\textbf{Science Cavity Mode Size. (left)} The waist size of the cavity is plotted as the total length of the cavity is scaled in proportion to the lower arm length. The nearly three-fold degenerate point we operate at is shown by the vertical line. \textbf{(right)} A plot of the mode size as a function of the distance along the optical axis throughout the cavity, including the small primary 19~$\mu$m waist in the lower arm (orange), where the atoms are placed, along with a larger 90~$\mu$m waist in the upper arm (green). The four cavity mirrors are labelled between the arms of the cavity with their radii-of-curvature in mm. The mode size is periodic in the cavity length, which is approximately 120~mm.}
	\label{Appx:SuppInfo-Fig:780NPCmodesize}
\end{figure}

\begin{figure*}
	\centering
	\includegraphics[width=\textwidth]{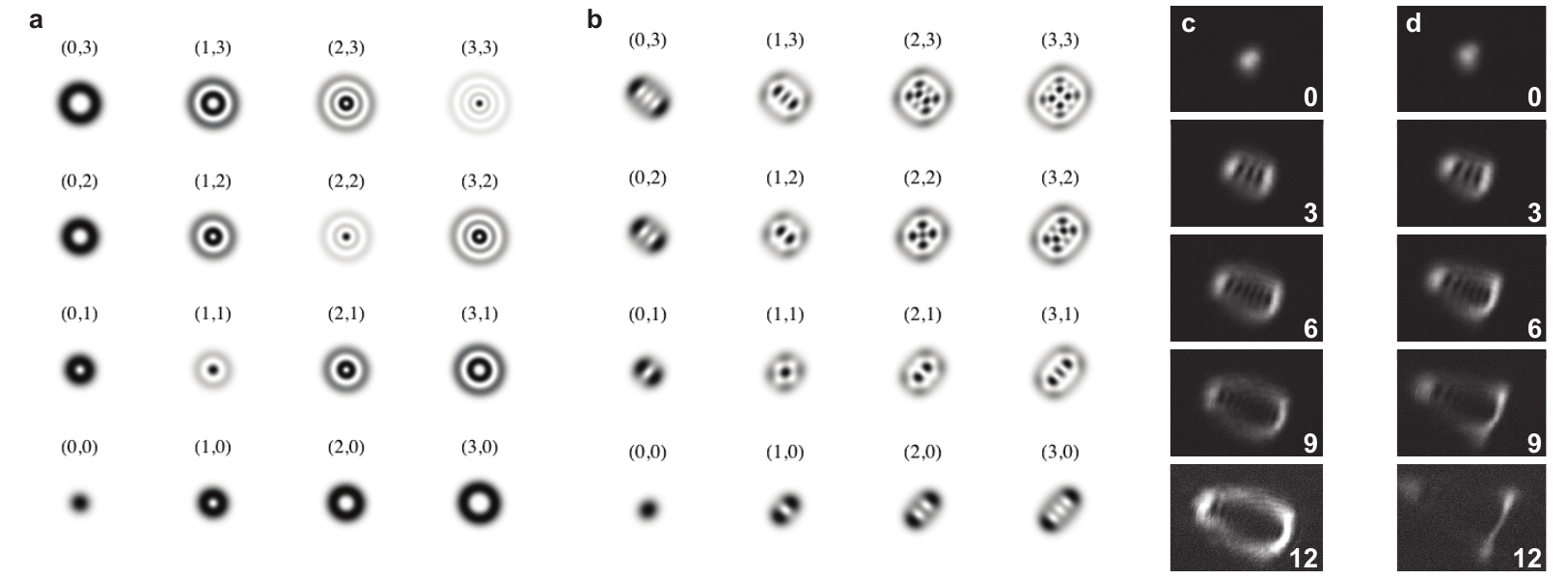}
	\caption{\textbf{Comparison of Mode Profiles. (a)} The cavity modes are expected to be near-perfect Laguerre Gaussian modes at the lower waist. \textbf{(b)} Significant astigmatism causes the cavity modes at the upper waist to have significant Hermite Gaussian character. \textbf{(c)} Away from degeneracy (30 MHz splitting between modes), the science cavity's eigenmodes are as expected; however \textbf{(d)} as we approach degeneracy (13 MHz splitting) the modes couple significantly, as is evident from the strong threefold symmetry that develops. The modes also become lossy, indicating that the coupling is from long range aberration rather than local disorder.}
	\label{Appx:SuppInfo-Fig:780modecomparison}
	\end{figure*}

	\subsection{Cavity details}
	\label{SI:Cavity}
	
This experiment utilizes two crossed nonplanar cavities; a primary `science' cavity for 780 nm photons and a build-up cavity for 480 nm photons. It is necessary to use a separate build-up cavity because the $480$~nm control field needs to cover an area larger than that spanned by the $780$~nm modes used in any experiment. To maintain sufficient control field Rabi coupling over the first five degenerate modes would then require $\sim10$~W of power in single pass. As this is significantly beyond maximum available laser power, we introduced a build-up cavity for the Rydberg field. This cavity is nonplanar to ensure circularly polarized modes, maximizing the Rabi coupling; it must also be a running wave cavity to avoid having nodes of the Rydberg field inside the atom cloud. 

To reduce noise and promote stability, both cavities' mirrors and piezos are mounted in a monolithic nonmagnetic steel structure. The science cavity is a 4 mirror, 3-fold degenerate nonplanar cavity modeled off of our previous work using such cavities without cold atoms present~\cite{Schine2016, Schine2018}. There were two primary design challenges in making these nonplanar cavities compatible with the atomic physics setup. First we needed a smaller waist size than we had previously achieved with our degenerate nonplanar cavities, since this cavity needed to be compatible with Rydberg mediated photon-photon interactions. This is a particular challenge because we also require Laguerre-Gaussian eigenmodes, and the standard techniques for reducing the waist such as reducing the radius-of-curvature of the cavity mirrors increases astigmatism, which then favors Hermite-Gaussian modes. While we previously explored single and two mode blockade physics in a cavity with a 14 $\mu$m waist~\cite{Jia2018b, Clark2019}, we relaxed the waist size requirement to $\sim 20$ $\mu$m and increased the target Rydberg level accordingly. This target was then achievable by using convex, $-75$ mm radius-of-curvature mirrors in the upper arm along with standard $25$ mm radius-of-curvature mirrors in the lower arm Fig.~\ref{Appx:SuppInfo-Fig:780NPCmodesize}.

The second design challenge for this cavity concerned reaching threefold degeneracy with the cavity in vacuum. Prior degenerate cavities were mounted in two halves separated by a micrometer stage. Adjusting this stage brought the modes into degeneracy. Since this technique is incompatible with vacuum, and the bake-out of the chamber was likely to change the cavity alignment slightly, we introduced a slow long throw piezo behind one of the cavity mirrors, providing a 63~$\mu$m free stroke, corresponding to an approximately 6\% change in the transverse mode spacing. Thus after aligning and gluing the cavity outside the vacuum chamber (and taking into account the index of refraction of air) and then baking out the chamber, degeneracy was within range of the long throw piezo.

\begin{figure*}[t]
	\centering
	\includegraphics[width=\textwidth]{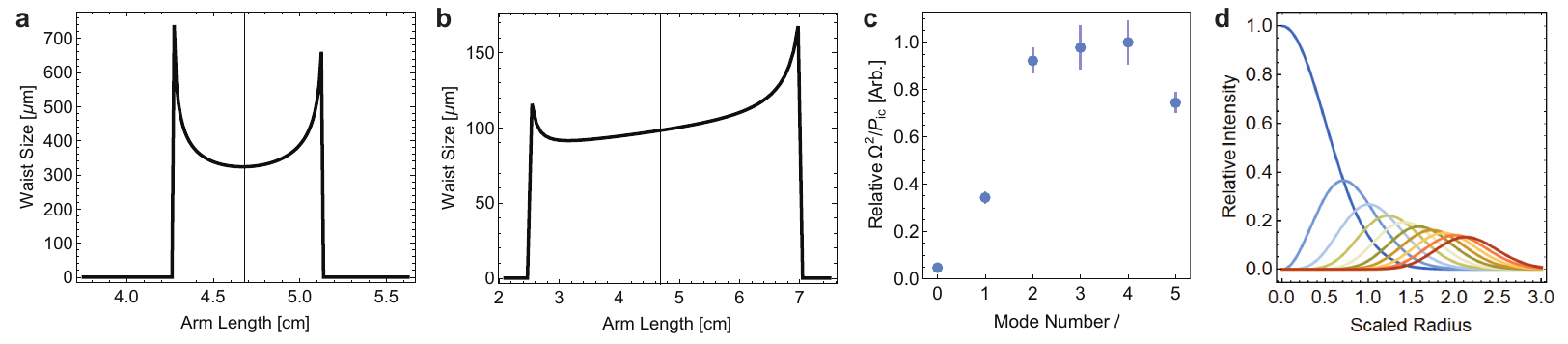}
	\caption{\textbf{Buildup Cavity Modes and Relative Alignment. (a)} The originally intended build-up cavity mode size was more than three times larger than \textbf{(b)} the mode size of the cavity after rebuilding to include the new high transmitter. \textbf{(c)} We fit the measured control Rabi frequency from cavity Rydberg EIT transmission at 40S and plot the ratio of that frequency to the intracavity circulating power versus the LG mode injected into the build-up cavity, finding a uniform factor of $\sim20$ improvement for $l = 2~,~3,~$or~$4$ compared to $l=0$. \textbf{(d)} This approximately uniform improvement indicates a relative shift between the mode centers of the two cavities by around 1.4 build-up cavity waists (140 $\mu$m). Each curve shows the radial intensity profile of a Laguerre-Gauss mode.}
	\label{Appx:SuppInfo-Fig:480CavInfo}
\end{figure*}

Otherwise, this cavity is fairly standard. The two convex mirrors are coated to outcouple light, with 99.91(1)\% reflection, while the lower mirrors are HR coated, so that there are effectively only two ports in this cavity. Similar coatings at $1560$~nm provide for convenient cavity length stabilization, enabled by a short-throw ring piezo stack actuating a cavity mirror. After construction and installation, we achieve a finesse of 1900(50) with a free spectral range of $2500(1)$ MHz at our operating point near, but not at, mode-degeneracy. 

The presence of significant astigmatism causes several issues in this cavity. First it makes the expected eigenmodes not Laguerre Gaussian in the upper waist. These modes are intermediate between Laguerre Gaussian and Hermite Gaussian and are connected to pure Laguerre Gaussian modes through an astigmatic mode converter ~\cite{beijersbergen1993astigmatic}. This does not affect the physics at the lower waist, at which the cavity modes are nearly-pure Laguerre Gaussian modes. Since we inject light into the cavity through the upper mirrors, we select the desired mode by programming the corresponding upper waist profile onto a digital micromirror device \cite{zupancic2016ultra}. The modes not being pure Laguerre Gaussian modes results in coupling between modes in our degenerate manifold, causing distortion and dramatically increased loss near degeneracy. Although the long throw piezo can move the cavity into degeneracy for every third mode in order to form a Landau level for cavity photons, the increased loss caused by astigmatism makes this useless. Thus we operate with Landau level modes split out by around 70 MHz (using both the long throw piezo and a heating wire), and make up the difference with Floquet modulation as discussed in Methods~\ref{methods:FloquetScheme}.

\begin{figure*}
	\centering
	\includegraphics[width=\textwidth]{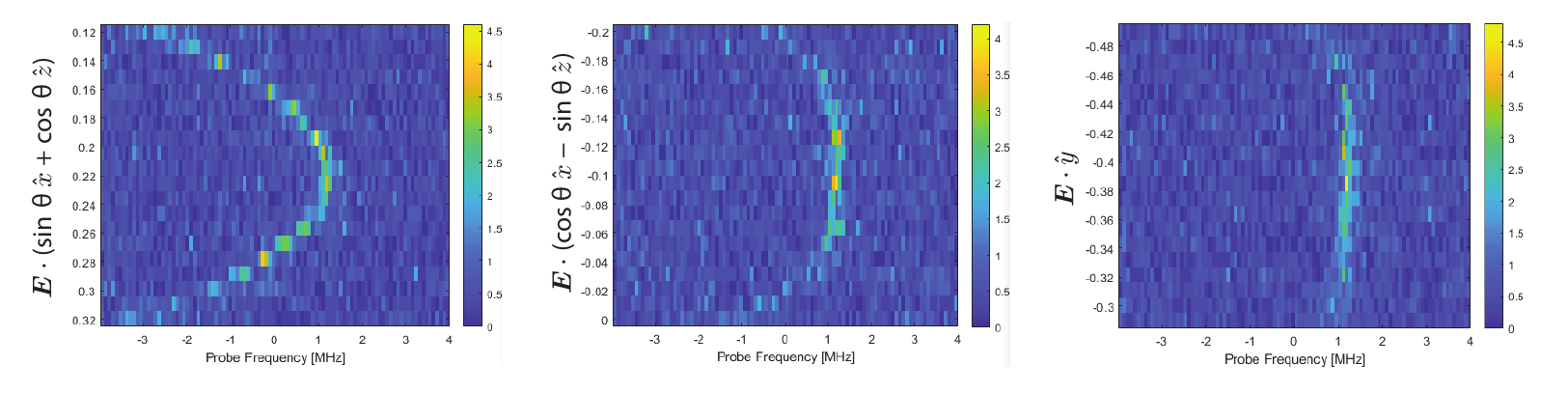}
	\caption{\textbf{Electric Field Scans.} The transmission rate (colorbar) for Rydberg-like dark polaritons is plotted versus probe frequency and applied electric field (units of V/dm). We scan the electric field $\mathbf{E}$ along three axes, as indicated on the individual plots, which are rotated around $\hat{y}$ from the typical coordinate axes used in this work by the angle $\theta\approx30^\circ$. To maintain an optically pumped sample, we apply a 5~G magnetic field along the blue cavity axis, which results in the large sensitivity in the leftmost plot, while suppressing the electric polarizability along the orthogonal axes observed in the other two plots. Strong orthogonal electric fields break the $m_j$ splitting along the magnetic field axis, cause coupling to many Rydberg states, and dramatically broaden and shrink the dark polariton resonance.}
	\label{Appx:SuppInfo-Fig:ElectricFieldScans}
\end{figure*}

Relative alignment between the two cavities is critical. Since the alignment was performed outside of vacuum before bakeout, we anticipated significant drift between the cavity modes. As such, we aimed to produce a very large waisted build-up cavity ($\sim320$~$\mu$m), with a corresponding increase in desired finesse. The increased mode area along with a desire for significantly higher blue Rabi frequency (to open up the possibility of increasing the atomic density) led to a mirror coating order specifying $99.97\%$ reflectors for a finesse 10,000 cavity. In fact, the mirrors we received were $99.985\%$ \textit{loss dominated} reflectors. This higher than expected loss at $480$ nm made these mirrors useless for the build-up cavity. Rather than wait the several months required to replace these mirrors, we instead used readily available mirrors with 1.5~m radius-of-curvature and a reflectivity of $97\%$ at $480$~nm. Rebuilding the cavity to include one of these mirrors produced a finesse 190 single ended cavity with a waist size of $98$~$\mu$m at the location of the atoms, providing enough intensity build-up to achieve a sufficient control Rabi frequency over the first few degenerate modes of the science cavity.

\begin{figure}
	\centering
	\includegraphics[width=\columnwidth]{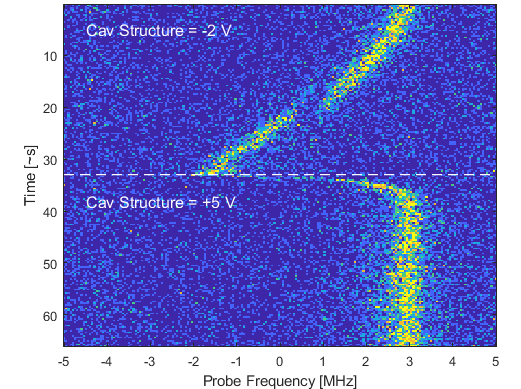}
	\caption{\textbf{Stray Charge Mitigation.} The transmission spectrum near the dark polariton resonance over time; blue indicates low transmission, yellow indicates high transmission. Illumination of the cavity structure with UV light can change the build up of stray charges that cause electric field drift and thus the drift of the dark polariton resonance. Furthermore, the simultaneous application of a voltage on the cavity structure dramatically changes the behavior of mobile charges. A substantial negative voltage appears to increase the drift rate of the electric field (above the white dashed line), while the application of a substantial positive voltage causes the field to rapidly equilibrate and strongly suppresses further drift; this suppression stabilizes the dark polariton resonance and provides the best conditions for our experiments (below the white dashed line).}
	\label{fig:SI_ChargeBuildupUV}
\end{figure}

Before bakeout, proper alignment between the two cavities was ensured by moving a thin wire cross on a translation stage and seeing both cavity modes spoil simultaneously. After installation and bakeout, there remained significant concern about the relative alignment between the cavity modes. Initial EIT signals showed a very weak control Rabi frequency. Coupling into higher order LG modes of the build-up cavity resulted in a much greater control Rabi frequency, normalized to the intracavity circulating power (See Fig.~\ref{Appx:SuppInfo-Fig:480CavInfo}c). Comparison with the relative intensities of the various higher order LG modes as a function of radius (Fig.~\ref{Appx:SuppInfo-Fig:480CavInfo}d) indicates that the improvement in overlap between the higher order build-up cavity modes and the science cavity mode arose from an $\sim140$~$\mu$m shift. 

In order to drive a higher order LG mode, we convert a large Gaussian beam into an LG$_{2,0}$ via a vortex half-wave retarder with $>80\%$ efficiency. This, combined with the significant $\sim 80\%$ peak intensity reduction still leaves a factor of 5 build-up in the circulating power.

\subsection{Electric field management}
\label{SI:ElectricField}

Special care was taken to engineer the electric field environment experienced by the Rydberg atoms. Most importantly, all surfaces, either metal or dielectric are kept at least 10 mm from the atoms, and line-of-sight between the atoms and dielectric surfaces is minimized to just the lower cavity mirrors. Piezos are shielded by grounded metal and are driven so that their surfaces that face the atoms are held at ground. In order to cancel the remaining background electric field $\mathbf{E}_{bkg}$ experienced by the atoms, the structure also supports eight electrodes which can apply additional electric fields and gradients at the location of the atomic cloud. The cavity structure itself is electrically isolated and controlled, acting as a ninth electrode. Typical scans of the dark polariton spectrum as a function of this applied electric field $\mathbf{E}$ are shown in Fig.~\ref{Appx:SuppInfo-Fig:ElectricFieldScans}. These scans enable us to identify the applied field which is necessary to achieve a net electric field $\mathbf{E}_{net}=\mathbf{E}_{bkg}+\mathbf{E}=0$ at the location of the atoms.

We observed drift of the electric field which we attribute to Rubidium (or Rubidium ion) accumulation on the cavity structure and mirrors. We mitigate this issue by pulsing a $\sim 100$ mW UV flashlight centered around 365 nm while jumping the set voltage of the entire cavity structure to +10 $V$ while the MOT is loading (Fig.~\ref{fig:SI_ChargeBuildupUV}). These passive mitigating factors along with active electric field control work in concert to enable our experiments with the $111D_{5/2}$ Rydberg state with negligible broadening due to electric field inhomogeneity and hours-long stability.

	\section{Theory}

	\subsection{Collective atomic excitations}
	\label{SI:CollectiveExcitations}
	Even though thousands of atoms are present in the cavity, only a small subset of the possible atomic excitations are coupled to the cavity photons. In fact, each cavity mode couples to a unique collective $5P_{3/2}$ excitation, which is subsequently coupled by the Rydberg laser to a unique collective Rydberg excitation (see Fig.~\ref{fig:FloquetPolaritonsFull}b). These collective excitations are superposition states in which one excitation is shared by many atoms with a spatial waveform that is proportional to the field of the corresponding cavity mode. In particular, photons in the cavity mode with angular momentum $l$ (created by the operator $a^\dagger_l$) couple to the collective $5P_{3/2}$ excitation created by the operator,
    \begin{equation*}
    p_{l}^{\dagger}=\frac{1}{g_l}\sum_{m=1}^{N_{at}}g_{ml}\sigma_{m}^{eg},
    \end{equation*}
    where $g_{l}^{2}\equiv\sum_{m=1}^{N_{at}}|g_{ml}|^{2}$ is the total atom-photon coupling strength for that mode and the coupling of cavity photons to the $m$'th atom $g_{ml}=g_\mathrm{sing} \phi_{l}(\mathbf{x}_{m})/\phi_{0}(0)$ is proportional to the electric field $\phi_{l}$ (normalized such that $\int d^2\mathbf{x} |\phi_l(\mathbf{x})|^2=1$) at the position $\mathbf{x}_{m}$ of the atom as well as the peak single-atom coupling strength $g_\mathrm{sing}$. The operator $\sigma_{m}^{eg}$ excites atom $m$ from the ground state to the $5P_{3/2}$ state. Similarly, the corresponding collective Rydberg excitation is created by the operator, 
    \begin{equation*}
    r_{l}^{\dagger}=\frac{1}{g_l}\sum_{m=1}^{N_{at}}g_{ml}\sigma_{m}^{rg},
    \end{equation*}
    where the operator $\sigma_{m}^{rg}$ excites atom $m$ from the ground state to the Rydberg state. Because the orthonormal set of cavity modes couples one-to-one with the orthonormal set of collective atomic excitations, the atom-cavity coupling does not directly mix the cavity modes together. Therefore, only the Rydberg-Rydberg interactions cause photons to move between the modes. For more on the orthogonality of the collective excitations see SI~B2 of Ref.~\cite{Clark2019}.

	\subsection{Photonic and polaritonic states differ in the Floquet scheme}
	\label{SI:PhotonsVsPolaritons}
	When forming dark polaritons in a set of near-degenerate cavity modes whose bare frequencies are all close to that of the unmodulated atomic transition, the relationship between photons and polaritons is independent of the mode. For example, because the atom-cavity coupling $g_l$ is the same for each mode $l$, the dark-state rotation angle $\theta_l$ is the same for every mode, where the rotation angle satisfies $\tan(\theta_l)\equiv g_l/\Omega$ (see~\cite{Jia2016} and~SI~B5~of~Ref.~\cite{Clark2019}). A smaller ratio $g_l/\Omega$ increases the contribution $\cos^2(\theta_l)$ to the polaritons from the cavity photon and thus makes the polariton more ``photon-like''; a larger ratio increases the contribution $\sin^2(\theta_l)$ from the collective Rydberg state, making the polariton more ``Rydberg-like''~\cite{Jia2016}. Therefore, when using a set of nearly-degenerate bare cavity modes, all of the polaritons have the same fractional composition in terms of their photonic and Rydberg parts. 
	
	The Floquet scheme causes the dark-state rotation angle, and thus the polariton composition, to vary between different modes. The rotation angle varies because the so-called ``sideband'' features with $l=3,~\&~9$ have smaller atom-cavity coupling than the ``carrier'' with $l=6$. For the parameters used in this work, the sideband dark polaritons are approximately six times more photon-like than the carrier polaritons. 
	
	The second difference between polaritons and photons in the Floquet scheme arises from the complex phase of $g_l$ and is more subtle to understand. Before considering the phases of the couplings, it is helpful to use a concrete example of the effect of the phase on a superposition of angular momentum states. For concreteness, consider the collective Rydberg states $\ket{R_\pm}=\frac{1}{\sqrt{2}} \left(r^\dagger_3\pm r^\dagger_6\right)\ket{vac}$. These Rydberg excitations form angular standing waves because of the interference between the $l=3$ and $l=6$ components; but the probability wave peaks of $\ket{R_+}$ match the troughs of $\ket{R_-}$, and vice versa, because of the flipped phase of the superposition. When utilizing ordinary polaritons with a constant $g_l\equiv{g}$, $\ket{R_\pm}$ couple with cavity photonic states $\ket{a_\pm}=\frac{1}{\sqrt{2}} \left(a^\dagger_3\pm a^\dagger_6\right)\ket{vac}$, where the peaks of $\ket{a_+}$ ($\ket{a_-}$) are at the same locations as the peaks of $\ket{R_+}$ ($\ket{R_-}$). Similar considerations apply for states containing multiple excitations; for ordinary polaritons, the spatial structures of the Rydberg components and the photonic components are the same.

    In the Floquet scheme, the structure of the single photon state $\ket{\tilde{a}_\pm}\propto \left(a^\dagger_3 \pm \frac{g_3}{g_6}e^{2\pi{i}f_\mathrm{mod}t} a^\dagger_6\right) \ket{vac}$ which couples to $\ket{R_\pm}$ is different in three important ways. First, the magnitude $\left|\frac{g_3}{g_6}\right|$ of the ratio between the coupling strengths determines the relative weights of the parts of the photonic superposition. Second, the relative phase $\mathrm{arg}\left(\frac{g_3}{g_6}\right)$ of the couplings affects the relative phase of the photonic superposition. When using sinusoidal Floquet modulation, the phases satisfy $\mathrm{arg}\left(\frac{g_3}{g_6}\right)+\mathrm{arg}\left(\frac{g_9}{g_6}\right)=\pi$. Third, in the lab frame the phase of the superposition rotates at the modulation frequency $f_\mathrm{mod}$. As a result, the positions of peaks and troughs in the probability wave will rotate around the center of the cavity at frequency $f_\mathrm{mod}/3$, where the factor of three arises because of the angular momentum difference between the modes in the superposition (equivalently, the fact that the density wave has three peaks). This rotation can be viewed as the micromotion which is quite common in Floquet systems.
    
    In this work, the two-Rydberg component of the polaritonic state produced by interactions is approximately,
    \begin{equation*}
    \ket{R_L}\propto \left( \sqrt{\frac{21}{10}} \left|\frac{g_3 g_9}{g_6^2}\right| r^\dagger_3r^\dagger_9 - \frac{1}{\sqrt{2}} \left(r^\dagger_6\right)^2 \right) \ket{vac},
    \end{equation*}
    noting that the factor $1/\sqrt{2}$ simply cancels a factor from the bosonic creation operators $\left(r^\dagger_6\right)^2$. The factor $\sqrt{21/10}$ is explained in SI~\ref{SI:VarietiesOfLaughlinStates}. As a result of the Floquet scheme, this Rydberg state couples to a two-photon state,
    \begin{align*}
    \ket{\tilde{a}_A} &\propto \left( \sqrt{\frac{21}{10}} e^{i~\mathrm{arg}(g_3 g_9/g_6^2)} a^\dagger_3a^\dagger_9 - \frac{1}{\sqrt{2}} \left(a^\dagger_6\right)^2 \right) \ket{vac}\\
    & \propto  \left( \sqrt{\frac{21}{10}} a^\dagger_3a^\dagger_9 + \frac{1}{\sqrt{2}} \left(a^\dagger_6\right)^2 \right) \ket{vac}.
    \end{align*}
    Two key features of this photonic state are worth noting: First, because the rotation factors at $\pm f_\mathrm{mod}$ from the $l=3$ and $l=9$ photons in the first term on the right-hand side cancelled out, there is no time-dependence of the Laughlin state itself. Even though all photonic states in the Floquet scheme rotate at $f_\mathrm{mod}/3$, the rotation-invariance of the Laughlin state prevents it from acquiring time-dependence. However, the net phase factors from the sideband couplings relative to the carrier do not cancel out; they contribute a factor $e^{i~\mathrm{arg}(g_3 g_9/g_6^2)}=-1$ which causes the photonic superposition to have a plus sign rather than a minus sign. In Methods~\ref{methods:SpaceCorr} we discuss the use of the phase compensation cavity to convert the state $\ket{\tilde{a}_A}$ which exits the science cavity into the photonic Laughlin state,
    \begin{align*}
    \ket{\tilde{a}_L} &\propto \left( \sqrt{\frac{21}{10}} a^\dagger_3a^\dagger_9 - \frac{1}{\sqrt{2}} \left(a^\dagger_6\right)^2 \right) \ket{vac}.
    \end{align*}

    While rotation-invariance makes the Laughlin state $\ket{\tilde{a}_L}$ time-independent, the rotation induced by the Floquet scheme plays a crucial role when measuring spatial correlations. The off-center single mode fiber measures localized photons; when viewed in the angular momentum basis, the localized fiber mode has an annihilation operator corresponding to a superposition $a_f\propto c_3 a_3 + c_6 a_6 + c_9 a_9$ where $c_l$ is the coefficient characterizing the contribution from mode $l$. When the fiber is translated to the radius with peak density, the coefficients are approximately $c_3=0.53$, $c_6=0.71$, and $c_9=0.46$. Immediately after a photon from the Laughlin state is detected through the single mode fiber, the remaining photon is in the state,
    \begin{align*}
    \ket{\tilde{a}_{rem}} & \propto  a_f\ket{\tilde{a}_L} \\ & \propto \bigg( \sqrt{\frac{21}{10}} \left(c_9e^{-2\pi{i}f_\mathrm{mod}t}a_3^\dagger + c_3e^{2\pi{i}f_\mathrm{mod}t}a_9^\dagger\right) \\ & ~~~~~~ - \sqrt{2}c_6 a_6^\dagger \bigg) \ket{vac}.
    \end{align*}
    Because the remaining photon is left behind in a state with spatial structure which is not rotation-invariant, the probability distribution of the remaining photon rapidly rotates in space. Immediately after the first photon is detected, the remaining photon has no overlap with the fiber mode and will not be observed; just half of a rotation period later the likelihood of detecting the photon is maximized. In particular, the likelihood of observing the second photon through the fiber follows,
    \begin{equation*}
        \bra{\tilde{a}_{rem}} a_f^\dagger a_f \ket{\tilde{a}_{rem}} \propto \sin^4\left(2\pi f_\mathrm{mod} t\right).
    \end{equation*}

	\subsection{Floquet polaritons are protected from intracavity aberrations}
	\label{SI:FloquetProtectsAberrations}
	
	It is often technically challenging to form a degenerate manifold
	of cavity modes, because mirror defects and intracavity aberrations
	such as astigmatism can induce couplings between the modes which break
	their degeneracy. Our Floquet scheme makes it possible to form an
	effectively degenerate manifold of dark polaritons, in which the polaritons
	have the same quasi-energy, without making the bare cavity modes degenerate.
	Naively, it might seem that the Floquet polaritons should inherit
	all of the couplings from their Rydberg and cavity photonic parts,
	in which case the degeneracy would still be split by the intracavity
	aberrations. However, as we will demonstrate below, the aberration
	couplings between Floquet polaritons are strongly suppressed, precisely
	because the large energy separation between the bare cavity modes
	remains. 
	
	We next demonstrate these features more formally. The behavior of
	Floquet polaritons is best understood in the high frequency approximation 
	\cite{Eckardt2015}, as detailed in Ref.~\cite{Clark2019}.
	We begin with the time-dependent Hamiltonian of the system, 
	
	\begin{align*}
	H(t) & =\sum_{n}^{N_{cav}}\bigg(\delta_{c}^{n}a_{n}^{\dagger}a_{n}+\bar{\delta}_{e}p_{n}^{\dagger}p_{n}+\delta_{2}r_{n}^{\dagger}r_{n}.\\
	& +\sum_{m=-\infty}^{\infty}g_{m}^{n}e^{im\omega t}p_{n}a_{n}^{\dagger}+h.c.+\\
	& +\sum_{m=-\infty}^{\infty}\Omega_{-m}e^{-im\omega t}r_{n}p_{n}^{\dagger}+h.c.\bigg)\\
	& +\frac{1}{2}\sum_{nmpq}^{N_{cav}}U_{nmpq}r_{n}^{\dagger}r_{m}^{\dagger}r_{p}r_{q}\\
	& +\sum_{n,q\neq n}^{N_{cav}}D_{nq}a_{n}^{\dagger}a_{q}+h.c.
	\end{align*}
	The first line accounts for the relative energies of the $N_{cav}=3$
	cavity modes, with energy $\delta_{c}^{n}$ for the $n$'th cavity
	mode with annihilation operator $a_{n}$, time-averaged energy $\bar{\delta_{e}}$
	of the collective $5P_{3/2}$ states with annihilation operator $p_{n}$,
	and energy $\delta_{2}$ for the collective Rydberg states with annihilation
	operator $r_{n}$. The next line denotes the atom-cavity coupling
	$g_{m}^{n}$ in mode $n$ through Floquet band $m$; $\omega=2\pi f_\mathrm{mod}$
	is the modulation angular frequency. The third line denotes the Rydberg
	couplings $\Omega_{m}$via Floquet band $m$. The fourth line accounts
	for the Rydberg-Rydberg interactions with strength $U_{nmpq}$ between
	all possible combinations of modes. The final line represents intracavity
	aberrations, which couple between cavity modes $n$ and $q$ with
	strength $D_{nq}$. 
	
	The experimentally relevant case is the limit in which each cavity
	mode $n$ is near-detuned to a band $k_{n}$ of the 5P$_{3/2}$ state
	and the Rydberg coupling laser is also near resonant for driving $5P_{3/2}\rightarrow111D_{5/2}$
	via band $l$. Under these conditions, we can write $\delta_{c}^{n}\equiv k_{n}\omega+\epsilon_{c}^{n},\bar{\delta}_{e}\equiv\epsilon_{p},\delta_{2}\equiv l\omega+\epsilon_{r},$
	such that the quasienergies satisfy, 
	\[
	\epsilon_{c},\epsilon_{p},\epsilon_{r}\ll\omega.
	\]
	
	Transforming to the frame of this resonant coupling, 
	\begin{align*}
	a_{n} & \rightarrow e^{ik_{n}\omega t}a_{n},\\
	& r_{n}\rightarrow e^{il\omega t}r_{n},
	\end{align*}
	the Hamiltonian becomes,
	
	\begin{align*}
	H(t) & =\sum_{n}^{N_{cav}}\bigg(\epsilon_{c}^{n}a_{n}^{\dagger}a_{n}+\epsilon_{p}p_{n}^{\dagger}p_{n}+\epsilon_{r}r_{n}^{\dagger}r_{n}\\
	& +\sum_{m=-\infty}^{\infty}g_{m+k_{n}}^{n}e^{im\omega t}p_{n}a_{n}^{\dagger}+h.c.+\\
	& +\sum_{m=-\infty}^{\infty}\Omega_{-m-l}e^{-im\omega t}r_{n}p_{n}^{\dagger}+h.c.\bigg)\\
	& +\frac{1}{2}\sum_{nmpq}^{N_{cav}}U_{nmpq}r_{n}^{\dagger}r_{m}^{\dagger}r_{p}r_{q}\\
	& +\sum_{n,q\neq n}^{N_{cav}}D_{nq}a_{n}^{\dagger}a_{q}e^{i(k_{q}-k_{n})\omega t}+h.c.
	\end{align*}
	Since we have now transformed to this basis where the quasienergies
	are all similar, the dynamics of the system will be dominated by the
	coupling terms which are not rotating. 
	
	Two features of this Hamiltonian are worthy of particular attention.
	First, because all of the bare Rydberg states are degenerate, and
	therefore couple to the P-state through the same Floquet band, all
	of the Rydberg-Rydberg interaction terms remain resonant (not rotating)
	even after this transformation. In contrast, because the bare cavity
	modes were not degenerate, and each couples to the P-state through
	a different Floquet band, the aberration couplings between the cavity
	mode now have a rapidly rotating complex phase, indicating that they
	are off-resonant.
	
	At lowest order in the high frequency approximation, which is equivalent
	to the rotating wave approximation at this order, the effective Hamiltonian
	is just the average of the full, time-dependent Hamiltonian above,
	
	\[
	H_{F}^{(1)}=H_{0},
	\]
	where $H_{m}$ is the $m$'th Fourier component of the Hamiltonian,
	
	\[
	H_{m}=\frac{1}{T}\int_{0}^{T}e^{-im\omega t}\tilde{H}(t)=H_{-m}^{\dagger}
	\]
	
	\begin{align}
	H_{F}^{(1)} & =\sum_{n}^{N_{\mathrm{cav}}}\bigg(\epsilon_{c}^{n}a_{n}^{\dagger}a_{n}+\epsilon_{e}p_{n}^{\dagger}p_{n}+\epsilon_{r}r_{n}^{\dagger}r_{n}\nonumber \\
	& +g_{k_{n}}^{n}p_{n}a_{n}^{\dagger}+\Omega_{-l}r_{n}p_{n}^{\dagger}+h.c.\label{eqn:HeffMultimode-1}\\
	& +\frac{1}{2}\sum_{nmpq}^{N_{cav}}U_{nmpq}r_{n}^{\dagger}r_{m}^{\dagger}r_{p}r_{q}\bigg)\nonumber \\
	& +\sum_{n,q\neq n}^{N_{cav}}D_{nq}\delta_{k_{q},k_{n}}a_{n}^{\dagger}a_{q}
	\end{align}
	As expected, at this level of approximation, mirror disorder is only
	relevant if the bare cavity modes are degenerate, as determined by
	the Kronecker delta $\delta_{k_{q},k_{n}}$ in the final line which
	requires that the modes couple to the P-state through the same Floquet
	band. The Rydberg interactions all contribute fully at
	the usual level because the bare Rydberg states in the absence of modulation were already degenerate. 
	
	At next order we see the effects of mirror disorder between cavity
	modes which are not naively degenerate. The next order contribution
	to the effective Hamiltonian is,
	
	\[
	H_{F}^{(2)}=\sum_{m\neq0}\frac{H_{m}H_{-m}}{m\hbar\omega},
	\]
	where the relevant Fourier components are,
	
	\begin{align*}
	H_{m} & =\sum_{n}^{N_{cav}}\bigg(g_{-m+k_{n}}^{n}p_{n}^{\dagger}a_{n}+g_{m+k_{n}}^{n}p_{n}a_{n}^{\dagger}\\ &+\Omega_{-m-l}r_{n}^{\dagger}p_{n}+\Omega_{m-l}r_{n}p_{n}^{\dagger}\bigg)\\
	& +\sum_{n,q\neq n}^{N_{cav}}D_{nq}a_{n}^{\dagger}a_{q}\delta_{m,k_{q}-k_{n}}+D_{nq}a_{n}a_{q}^{\dagger}\delta_{-m,k_{q}-k_{n}}.
	\end{align*}
	At this order, the effective Hamiltonian gains a variety of terms
	with strengths $D^{2}/\omega$, $Dg/\omega$, or $D\Omega/\omega$.
	As long as all of the coupling strengths $g,\Omega,D\ll\omega$ are
	small compared to the modulation frequency, these terms will be negligible.
	Overall, this treatment demonstrates how the Floquet scheme enables
	us to create degenerate manifolds of dark polaritons while strongly
	suppressing the broadening that intracavity aberrations would cause
	in a degenerate manifold of bare cavity modes.

	\subsection{Many-body spectrum}
	\label{SI:ManyBodySpectrum}
	The many-body spectrum for two excitations in the three modes accessible in this paper (Fig.~3b of the main text) is calculated as follows. When the three modes have been made degenerate with energy per particle $E_{pol}$, the energies of multi-particle states are only differentiated by the interaction Hamiltonian,
	\begin{equation}
	H_{\mathrm{int}}=\frac{U}{2}\int d^{2}z\psi^{\dagger}(z)\psi^{\dagger}(z)\psi(z)\psi(z),
	\end{equation}
	where $\psi(z)$ is the field operator and we have made the approximation of contact interactions with strength $U$. The approximation of contact interactions is best justified when the Rydberg blockade radius is small compared to the cavity mode waist. However, even when the blockade radius and mode waist are comparable, the spectrum remains qualitatively similar. Note that the interaction strength $U$ typically has a large imaginary component~\cite{Georgakopoulos2018}, but we depicted a larger real than imaginary component in Fig.~3a~\&~b to make the visualization clear.
	
	Projecting the interaction Hamiltonian into the basis of our degenerate Landau level yields,
	\begin{equation}
	 H_{\mathrm{int}}=\frac{U}{2}\sum_{ijkl\in\{3,6,9\}}\beta_{ijkl}a_{i}^{\dagger}a_{j}^{\dagger}a_{k}a_{l},
	 \end{equation}
	 where the interaction energies among modes are determined by the overlap integrals,
	 \begin{equation}
	 \beta_{ijkl}=\int d^{2}z\phi_{i}^{*}(z)\phi_{j}^{*}(z)\phi_{k}(z)\phi_{l}(z),
	 \end{equation}
	 and the wavefunction for the mode with angular momentum $\hbar l$ is,
	 \begin{equation}
	 \phi_{l}(z)=\frac{1}{\sqrt{2^{l}l!}}z^{l}e^{-|z|^{2}/4}.
	 \end{equation}
	Diagonalizing this interaction Hamiltonian within the two-particle manifold yields the energy spectrum of two-particle states in our system, including the Laughlin state with zero interaction energy.

	\subsection{Varieties of Two-Particle Laughlin States}
	\label{SI:VarietiesOfLaughlinStates}
	Two photons in any three evenly spaced angular momentum modes can form a Laughlin state. While these Laughlin states differ in the exact form of their wavefunctions, they all enable the two particles to minimize their interaction energy while remaining in the lowest Landau level. The two particle Laughlin states in which we are interested can be expressed in real space as,
	\begin{equation}
	\psi_{L}(z_{1},z_{2};m,n)=N_{mn}z_{1}^{m}z_{2}^{m}(z_{1}^{n}-z_{2}^{n})^{2}e^{-(|z_1|^2+|z_2|^2)/4},
	\label{eqn:psiL_spatial}
	\end{equation}
	\noindent where $z_j\equiv x_j + iy_j$ represents the position of particle $j$ in units of the magnetic length, $m~\&~n$ are positive integers, and $N_{mn}$ is an overall normalization factor ensuring $\int\int d^2z_1 d^2z_2 |\psi_L|^2 = 1$. Note that these states all share the property that particles avoid each other, since $\psi_{L}(z_{1}=z_{2};m,n)=0$. 
	
	The two particle Laughlin states $\psi_{L}(z_{1},z_{2};m,n)$ can be composed from the three angular momentum modes with $l=m,~m+n,$~and~$m+2n$. To explicitly perform the transformation to the angular momentum basis we can expand the quadratic factor, yielding, 
	\begin{multline}
	\psi_{L}(z_{1},z_{2};m,n)=N_{mn} e^{-(|z_1|^2+|z_2|^2)/4}\\
	\times \left(z_{1}^{m}z_{2}^{m+2n}+z_{1}^{m+2n}z_{2}^{m}-2z_{1}^{m+n}z_{2}^{m+n}\right).
	\label{eqn:psiLexpanded}
	\end{multline}
	\noindent Comparing this form to the single particle angular momentum modes,
	\begin{equation}
	\phi_{l}(z)=\frac{1}{\sqrt{2^{l}l!}}z^{l}e^{-|z|^{2}/4},
	\end{equation}
	\noindent reveals that the first two terms on the right-hand side of Eq.~\ref{eqn:psiLexpanded} correspond to the properly symmetrized two-particle angular momentum state $\phi_{l_1, l_2}$ for particles with angular momenta $l_1=m$ and $l_2=m+2n$. The last term on the right-hand side of Eq.~\ref{eqn:psiLexpanded} corresponds to $\phi_{m+n, m+n}$. Accounting for normalization factors of each angular momentum state, we obtain,
	\begin{multline}
	\psi_{L}(z_{1},z_{2};m,n)=N_{mn} e^{-(|z_1|^2+|z_2|^2)/4}\\
    \times\big(\sqrt{2}\sqrt{2^{2n+m}(2n+m)!}\sqrt{2^{m}(m)!}\phi_{2n+m,m}(z_{1},z_{2})\\
	-2^{n+m+1}(n+m)!\phi_{n+m,n+m}(z_{1},z_{2})\big),
	\end{multline}
	or approximately in the notation of the main text (where $\ket{uv}\equiv\ket{u,v}$ means a two-particle state with the individual particles possessing angular momenta $u\hbar$ and $v\hbar$ about the origin),
	\begin{multline}
    \ket{L}_{mn}=\\
	N_{mn}\big(\sqrt{2}\sqrt{2^{2n+m}(2n+m)!}\sqrt{2^{m}(m)!}\ket{m,m+2n}\\
	-2^{n+m+1}(n+m)!\ket{m+n,m+n}\big).
	\end{multline}
	Therefore, in each Laughlin state, the ratio $\alpha_{mn}$ between the populations in $\ket{m,m+2n}$ and $\ket{m+n,m+n}$ is,
	\begin{equation}
	\alpha_{mn}=\frac{(2n+m)!m!}{2(n+m)!^{2}}.
	\end{equation}
	For the Laughlin state with $m=n=3$ used in this work, we find $\alpha_{mn}=\frac{21}{10}=2.1$, as reported in the main text. Note that, for each Laughlin state $\ket{L}_{mn}$, there is a corresponding ``anti-Laughlin'' state,
	\begin{multline}
	\ket{AL}_{mn}=\\
	N_{mn}\big(2^{n+m+1}(n+m)!\ket{m,m+2n}\\
	+\sqrt{2}\sqrt{2^{2n+m}(2n+m)!}\sqrt{2^{m}(m)!}\ket{m+n,m+n}\big),
	\end{multline}
	which exhibits spatial bunching instead of anti-bunching. 
	
	Physically, one can think of the choice of $n$ as splitting the ordinary Landau level (containing every angular momentum mode $l=0,1,2...$) into $n$ separate Landau levels existing on cones with a spatial curvature of $R(x,y)=4\pi\left(1-\frac{1}{n}\right)\delta^{(2)}(x,y)$ localized at the cone tip, where $\delta^{(2)}(x,y)$ is the two-dimensional Dirac delta function~\cite{Schine2016, wu2017fractional, Schine2018}. The $n$ cones with the same curvature correspond to different choices of the lowest angular momentum mode $m$, which determines the effective magnetic flux $\frac{m~\mathrm{mod}~n}{n}\Phi_0$ threaded through the cone tip, where $\Phi_0$ is the magnetic flux quantum. Moreover, when $m$ exceeds $n$, the Laughlin state has $\mathrm{Floor}(m/n)$ quasi-holes pinned at the origin, where $\mathrm{Floor}(x)$ denotes the largest integer less than or equal to $x$.

    Interestingly, the azimuthal correlations $g^{(2)}_{mn}(\phi)\propto|\psi_L(z_{1}=z_{2}e^{i\phi};m,n)|^2$ are independent of $m$ and have a very simple dependence on $n$. Using the form of the wavefunction in Eq.~\ref{eqn:psiL_spatial}, it is straightforward to find that,
	\begin{equation}
	g^{(2)}(\phi)_{mn}\propto\sin^4\left(\frac{n}{2}\phi\right).
	\end{equation}
	Moreover, when viewed in terms of the angle $\phi_\mathrm{cone}=n\phi$ around the cone tip, the azimuthal correlation function is entirely independent of $m$ and $n$. 
	
	Since these Laughlin states are all closely related physically, our choice of $m=n=3$ was made for technical reasons. First, we chose to make only every third angular momentum state degenerate to protect the Landau level from intracavity astigmatism (see Ref.~\cite{Schine2016}). In the end, because we use Floquet polaritons, this choice was likely not necessary (see SI~\ref{SI:FloquetProtectsAberrations}); however, the length of our cavity is not sufficiently tunable to bring any other set of angular momentum modes near enough to degeneracy to be convenient for our Floquet scheme. Second, we choose $m=3$, making $l=3$ our lowest angular momentum mode rather than $l=0$, because it yields $\alpha_{33}=2.1$ much smaller than $\alpha_{03}=10$. Smaller $\alpha_{mn}$ increases the contribution of $\ket{m+n,m+n}$ to the Laughlin state, improving the coupling of our coherent probe on $l=m+n$ to the Laughlin state relative to the anti-Laughlin state.

\end{document}